\documentclass[preprintnumbers, floatfix, preprintnumbers, letterpaper, superscriptaddress,nofootinbib]{revtex4-2}
\usepackage{graphicx}
\usepackage{microtype}
\usepackage{amsmath}
\usepackage{amssymb}
\usepackage{subfigure}
\usepackage{hyperref}
\usepackage{url}
\usepackage{xcolor}
\usepackage{color} 
\usepackage{mathrsfs}
\usepackage{calrsfs}
\usepackage{amsfonts}
\usepackage{lipsum}
\usepackage{eufrak}
\usepackage{tabularx}
\usepackage{eucal}
\usepackage{latexsym}
\usepackage{ragged2e}
\usepackage{epsfig}
\usepackage{textcomp}
\usepackage{adjustbox}
\usepackage{caption}
\usepackage{soul}
\usepackage{orcidlink}
\usepackage{float}
\DeclareCaptionJustification{justified}{\leftskip=0pt \rightskip=0pt \parfillskip=0pt plus 1fil}
\definecolor{vividviolet}{rgb}{0.62, 0.0, 1.0}
\definecolor{amaranth}{rgb}{0.9, 0.17, 0.31}
\definecolor{palatinateblue}{rgb}{0.15, 0.23, 0.89}
\definecolor{brightpink}{rgb}{1.0, 0.0, 0.5}
\definecolor{cornflowerblue}{rgb}{0.39, 0.58, 0.93}
\definecolor{deepcarminepink}{rgb}{0.94, 0.19, 0.22}
\definecolor{radicalred}{rgb}{1.0, 0.21, 0.37}

\hypersetup{ linktoc=all,
	colorlinks, linkcolor={palatinateblue},
	citecolor={brightpink}, urlcolor={amaranth}
}

\graphicspath{{Images/}}

\def\sideremark#1{\ifvmode\leavevmode\fi\vadjust{\vbox to0pt{\vss
			\hbox to 0pt{\hskip\hsize\hskip1em
				\vbox{\hsize1.3cm\tiny\raggedright\pretolerance10000
					\noindent #1\hfill}\hss}\vbox to8pt{\vfil}\vss}}}%
\def\beq{\begin{equation}}
\def\eeq{\end{equation}}
\newcommand{\be}{\begin{equation}}
\newcommand{\ee}{\end{equation}}
\newcommand{\ba}{\begin{eqnarray}}
\newcommand{\ea}{\end{eqnarray}}

\begin{document}

\title{Reproducing $\Lambda$CDM-like Solutions in $f(Q)$ Gravity: A Comprehensive Study Across All Connection Branches}

\author{Saikat Chakraborty\orcidlink{0000-0002-5472-304X}}
\email{saikat.ch@nu.ac.th}
\email{48674087@nwu.ac.za}
\affiliation{The Institute for Fundamental Study, Naresuan University, Phitsanulok 65000, Thailand}
\affiliation{Center for Space Research, North-West University, Potchefstroom 2520, South Africa}

\author{Jibitesh Dutta\orcidlink{0000-0002-6097-454X}}
\email{jibitesh@nehu.ac.in}
\affiliation{Mathematics Division, Department of Basic Sciences and Social Sciences, North Eastern Hill University, Shillong, Meghalaya 793022, India}
\affiliation{Inter University Centre for Astronomy and Astrophysics, Pune 411 007, India}

\author{Daniele Gregoris\orcidlink{0000-0002-0448-3447}}
\email{danielegregoris@libero.it}
\affiliation{School of Science, Jiangsu University of Science and Technology, Zhenjiang 212100, China}

\author{Khamphee Karwan}
\email{khampheek@nu.ac.th}
\thanks{Corresponding author}
\affiliation{The Institute for Fundamental Study, Naresuan University, Phitsanulok 65000, Thailand}

\author{Wompherdeiki Khyllep\orcidlink{0000-0003-3930-4231}}
\email{sjwomkhyllep@gmail.com}
\affiliation{Department of Mathematics, St. Anthony’s College, Shillong, Meghalaya 793001, India}

\begin{abstract} 
	Given the remarkable success of the $\Lambda$CDM model in fitting various cosmological observations, a pertinent question in assessing the phenomenological viability of modified gravity theories is whether they can reproduce an exactly $\Lambda$CDM-like cosmic background evolution. In this paper, we address this question in the context of $f(Q)$ gravity, where $Q$ denotes the nonmetricity scalar. It is known that there are three possible symmetric teleparallel connection branches that respect the cosmological principles of spatial homogeneity, isotropy, and global spatial flatness. By enforcing a $\Lambda$CDM-like background evolution via the cosmographic condition $j(z) = 1$, where $j$ is the jerk parameter, we reconstruct the $\Lambda$CDM-mimicking $f(Q)$ theory for each of the three possible connection branches. For the first connection branch, also known as the ``coincident gauge'' in cosmology, we recover the previously known result that a theory of the form $f(Q) = -2\Lambda + \alpha Q + \beta\sqrt{-Q}$  can exactly reproduce a $\Lambda$CDM-like cosmic evolution. Furthermore, we establish that the stability of the $\Lambda$CDM-like cosmic solution within this reconstructed $f(Q)$, as well as the robustness of the reconstructed $f(Q)$ form with respect to small errors in the astrophysical measurements of the jerk parameter. For the second connection branch, we analytically reconstruct the $\Lambda$CDM-mimicking $f(Q)$ to be of the form $f(Q) = -2\Lambda + \alpha Q - \beta Q^2$. For the third connection branch, we could decouple the evolution equation for the dynamical connection function, which enabled us to perform a numerical reconstruction. Our analysis proves that, at least at the background level, it is possible to obtain $\Lambda$CDM-mimicking $f(Q)$ models for all the three possible connection branches.  These models effectively  reduce to (STE)GR in the past while possessing a positive effective gravitational coupling throughout.
\end{abstract}

\maketitle

\section{Introduction} 

General Relativity (GR), based on the curvature of spacetime, has been fundamental in understanding the physics of gravitational interaction and the dynamics of the universe's large-scale structure. Its achievements encapsulated within the $\Lambda$CDM cosmological model include the successful description of the cosmic microwave background (CMB), baryonic acoustic oscillations (BAO), and the accelerating expansion of the universe \cite{SupernovaSearchTeam:1998fmf, SupernovaCosmologyProject:1998vns}. Still, specific unresolved theoretical problems such as the cosmological constant problem \cite{Weinberg:1988cp} and observational tensions such as the Hubble tension \cite{Verde:2019ivm} provide reasons to study alternatives to GR.

The geometric foundation of GR is, however, part of a broader ``Trinity of Gravity'' that also includes two equivalent formulations of GR: the teleparallel equivalent of GR (TEGR) and the symmetric teleparallel equivalent of GR (STEGR) \cite{BeltranJimenez:2019esp}. In the TEGR framework, gravity is described through the torsion of the spacetime \cite{Aldrovandi:2015wfa}, while in STEGR, it is interpreted in terms of nonmetricity of the spacetime \cite{BeltranJimenez:2017tkd}. These alternative formulations of non-Riemannian geometry provide fertile grounds for developing modified gravity theories based on arbitrary scalar functions of their respective scalar quantities, such as $f(T)$ or $f(Q)$, where $T$ and $Q$ represent the torsion and nonmetricity scalars, respectively \cite{Boehmer:2023fyl,Boehmer:2021aji,Boehmer:2022wln}.

There has been tremendous interest in $f(Q)$ gravity in recent years due to its exceptional geometric interpretation without using curvature or torsion. However, we point out that recently there has been a discussion about potential presence of ghost and strong coupling issues in $f(Q)$ gravity \cite{Gomes:2023tur}, and whether such ghost degrees of freedom can actually propagate \cite{Hu:2023gui}. In fact, similar issues also showed up in the study of $f(T)$ gravity, where certain degrees of freedom lose their kinetic terms on the FLRW background \cite{BeltranJimenez:2020fvy}. It was pointed out in \cite{Heisenberg:2023wgk} that a direct coupling of the matter field to the connection or a nonminimal coupling between the matter and the geometry sector might alter the situation leading to a healthy theory. A class of ghost free theories in symmetric teleparallel geometry was recently investigated in \cite{Bello-Morales:2024vqk}. Strong coupling and ghost issues, as well as the question of the actual number of propagating degrees of freedom underscore the need for perturbative stability analysis and Hamiltonian analysis in $f(Q)$ theory. Investigating these issues are important from the theoretical point of view for establishing the viability of modifying gravity in the symmetric teleparallel framework. Indeed, this is an active area of research currently. 

Although we are aware of the above theoretical issues, this is not the main focus of our investigation here. Rather, our focus is to examine the phenomenological viability of modifying gravity in the symmetric teleparallel framework, along a line of similar works \cite{Anagnostopoulos:2021ydo,Atayde:2021pgb,Albuquerque:2022eac,Sahlu:2024pxk,Goncalves:2024sem,De:2022jvo}. The framework of symmetric teleparallelism gives promising insights into cosmological dynamics. $f(Q)$ gravity has proven to be compatible with the observational data at the level of the background and of the perturbations, such as CMB, supernovae, BAO, redshift-space distortions, growth measurements \cite{Lazkoz:2019sjl,Ayuso:2020dcu,Frusciante:2021sio}. In addition, it does not have problems with the constraints imposed by Big Bang Nucleosynthesis, which means it could successfully compete with other modified gravity theories \cite{Anagnostopoulos:2022gej}.

The $\Lambda$CDM model based on Einstein's GR remains a very good fit with available cosmological datasets at least according to Planck 2018 \cite{Planck:2018vyg}\footnote{The recent DESI data \cite{DESI:2023ytc} apparently favours an evolving dark energy, although there has been arguments that their conclusion is highly driven by the choice of the priors \cite{Cortes:2024lgw}.}\footnote{An interesting model was proposed in \cite{Anagnostopoulos:2022gej}: $f(Q)=Qe^{\lambda\frac{Q_0}{Q}}$. This model, in coincident gauge, although not having any $\Lambda$CDM-limit in the FLRW cosmology at all, was slightly preferred against the standard cosmological model in fitting various datasets  according to the information criteria.}. In the context of modified gravity theories, from a phenomenological point of view, an important question is whether a modified gravity theory can give rise to exactly the same kind of background evolution as the GR-based $\Lambda$CDM model. Such modified gravity theories will then be indistinguishable from the $\Lambda$CDM model at the background level, and typically one needs to go to the perturbation level to find any distinctive signature. This particular question has been addressed in the context of $f(R)$ gravity \cite{Dunsby:2010wg,He:2012rf,Fay:2007uy}, $F(R,G)$ and modified Gauss-Bonnet gravity \cite{Elizalde:2010jx} and coincident gauge $f(Q)$ gravity (i.e. considering only the trivial connection branch where $Q=-6H^2$) \cite{Gadbail:2022jco,Albuquerque:2022eac}. However, in general there are three different symmetric teleparallel connection branches respecting spatial flatness, homogeneity and isotropy \cite{Hohmann:2021ast}, which we denote later as $\Gamma_1,\,\Gamma_2,\,\Gamma_3$, with $\Gamma_2,\,\Gamma_3$ commonly being referred to in the literature as the so-called non-coincident gauges \cite{Paliathanasis:2023hqq}\footnote{There is a slight misnomer here. For any particular connection branch $\Gamma_i$, one can find a coordinate system in which the connection can be trivialized \cite{Hohmann:2021ast}. This particular coordinate system is the coincident gauge of the connection branch in question. It happens that the coincident gauge for the connection branch $\Gamma_1$ is the usual Cartesian coordinate system.}. In this paper, we reconstruct the $f(Q)$ gravity that reproduces exactly a $\Lambda$CDM-like background evolution for each of the three connection branches, recovering also the earlier reconstructed $f(Q)$ for $\Gamma_1$.

There has been some work on the cosmological reconstruction of $f(Q)$ gravity, primarily within the trivial connection branch $\Gamma_1$, i.e., the so-called coincident gauge. For instance, \cite{Capozziello:2022wgl} numerically reconstructed $f(Q)$ starting from a cosmographic series with Padé approximations, while \cite{Esposito:2021ect} presented reconstruction algorithms for Bianchi-I and FLRW cosmology. In \cite{Albuquerque:2022eac}, a designer approach was adopted to reconstruct $f(Q)$ corresponding to various dark energy equation of state parametrizations, and \cite{Nojiri:2024zab} explored cosmological reconstruction of both $f(Q)$ gravity and mimetic $f(Q)$ gravity. Similarly, \cite{Gadbail:2022jco} reconstructed $f(Q)$ gravity that reproduces a $\Lambda$CDM-like background evolution.

To our knowledge,  the paper \cite{Yang:2024tkw} is the first study to attempt reconstruction beyond the coincident gauge. Using the OHD dataset and assuming vanishing hypermomentum, the authors numerically reconstructed $f(Q)$ for the connection branch $\Gamma_2$ \cite[Eq.30]{Yang:2024tkw}. In the general case with nonvanishing hypermomentum, they numerically reconstructed the dynamical connection function for $\Gamma_2$ and $\Gamma_3$ by assuming particular forms of $f(Q)$.

In this paper, we perform the cosmological reconstruction of $f(Q)$ starting from the $\Lambda$CDM-like cosmic evolution for all three connection branches. For the connection branch $\Gamma_1$, we recover the results earlier obtained in \cite{Albuquerque:2022eac,Gadbail:2022jco}. For the connection branch $ \Gamma_2 $, we could perform an analytical reconstruction and obtain a functional form for the $ \Lambda $CDM-mimicking $f(Q)$. For the connection branch $\Gamma_3$, which we found to be the most complicated, we could obtain a decoupled evolution equation for the dynamical connection function, which ultimately enabled us for a numerical reconstruction of the $\Lambda$CDM-mimicking $f(Q)$.

The paper is organized as follows:  In sections \ref{sec:ss2}, \ref{sec:f(Q)}, we give some brief and self-contained exposures to the cosmographic condition corresponding to the $\Lambda$CDM-like cosmic evolution and $f(Q)$ gravity respectively. In section \ref{sec:FLRW_f(Q)}, we introduce the three connection branches of the FLRW cosmology in the symmetric teleparallel framework and write down the corresponding cosmological field equations in section \ref{sec:field_eqs}. The sections \ref{sec:recon_Gamma1}, \ref{sec:recon_Gamma2}, \ref{sec:recon_Gamma3} are dedicated to the reconstruction of the $\Lambda$CDM-mimicking $f(Q)$ for the connection branches $\Gamma_1$, $\Gamma_2$ and $\Gamma_3$ respectively. In each case, we identify the free parameters of the reconstructed $f(Q)$ model and possible bounds for them, as much as possible. We also show the evolution of the effective gravitational coupling in each case. For the case of $\Gamma_1$, we also investigate the stability of the $\Lambda$CDM-like cosmological solution in the reconstructed $f(Q)$ in the subsection \ref{subsec:stab_Gamma1} and the robustness of the reconstructed $f(Q)$ against astrophysical errors of the measurement of the jerk parameter in section \ref{subsec:rob_Gamma1}. Finally, we conclude in section \ref{sec:sum_disc} summarizing the key takeaways from our paper and adding some relevant discussions. Additionally, in appendix \ref{app:distinguish} we put a brief discussion on the possibility of distinctive signature of the $\Lambda$CDM-model and the $\Lambda$CDM-mimicking $f(Q)$ models at the perturbation level, in appendix \ref{app:stab_Gamma2} we comment on how one can go ahead to investigate the stability of the $\Lambda$CDM-like cosmological solution in the reconstructed $f(Q)$ for the case of $\Gamma_2$ and in appendix \ref{app:gamma_Gamma2} add some comments about the dynamical connection function for $\Gamma_2$.

Throughout the paper, we use the metric signature $(-,+,+,+)$ and the geometrized unit system $8\pi G=c=1$, $G$ being Newton's gravitational constant and $c$ is the speed of light in vacuum.


\section{Hubble rate from cosmographic reconstruction}\label{sec:ss2}

Cosmography is a model-independent approach to derive the universe's expansion history from astrophysical datasets, characterized by a hierarchy of parameters involving time derivatives of the scale factor $a(t)$. The first two terms of this hierarchy are the Hubble rate and deceleration function, respectively:
\beq
H:= \frac{\dot a}{a}\,, \qquad q:= - \frac{\ddot{a} a}{\dot a^2}=-1-\frac{\dot H}{H^2}\,,
\eeq
while the third one is the so-called jerk parameter \cite{naturej}:
\be\label{j-def}
j = \frac{\dddot{a}}{aH^3} = 1 + 3\frac{\dot{H}}{H^2} + \frac{\ddot{H}}{H^3}\,.
\ee
The cosmographic parameters characterize the `kinematics' of the universe, capturing information from the time derivatives of its scale factor. However, they do not provide  an understanding of its dynamics.
A relationship of the cosmographic parameters to the matter abundance parameters can be found only if a specific `model', as characterized by the Friedmann and Raychaudhuri equations , is specified \cite{Chakraborty:2022evc}. In other words, while the cosmographic parameters are completely kinematic, the particular relationship between the matter abundance parameter to the cosmographic parameters depends on the `dynamics'. For example, in the $\Lambda$CDM model, the matter abundance parameter is related to the cosmographic deceleration parameter as
\be 
\Omega_{m0}=\frac{2}{3}(1+q_0)\,,
\ee 
where the subscript `$0$' implies present day values. For alternative late time models, this relationship will change, as we will see later on in the text for $f(Q)$ gravity. It is worth mentioning that a different algebraic combinations of the time derivatives could also be considered as in statefinder diagnostic approach \cite{Sahni:2002fz}. 

The evolution of the cosmographic functions with respect to the redshift can be inferred in a model-independent (or non-parametric) manner directly from relevant datasets. Deviations from the a $\Lambda$CDM{\it like} evolution being constrained to be no more than 10 (per cent) \cite[Fig.11]{Bernal:2016gxb}-\cite[Fig.10]{Mukherjee:2020ytg}-\cite[Fig.4]{Jiang:2024xnu}-\cite{Pogosian:2021mcs}. The spatially flat $\Lambda$CDM model exhibits a very special cosmographic property of a constant jerk parameter $j(z)=1$ all along its evolution. A stout theory should not be fine-tuned to just a very specific numerical value of the jerk, but rather be applicable to all its astrophysical values  within an interval of observational uncertainty. We shall therefore set $j=1+\varepsilon=const.$ for ``small'' $\varepsilon$; the latter should nevertheless be regarded as the previously mentioned largest possible uncertainty in the observational estimate of the jerk function so that the evolution provided by different redshift parametrization would all fall within this interval.  This choice has been shown not to trouble the clustering properties of dark matter accounting for a consistent structure formation cosmic epoch \cite{Bardeen:1980kt}, but sill allowing for small departures from a $\Lambda$CDM-like evolution which may come in handy in  taming  current observational tensions \cite{Luongo:2013rba}.  By integrating (\ref{j-def}), we obtain \cite[Eq.(8)]{Amirhashchi:2018vmy}: 
\beq
\label{generalconstj}
H^2(z)= H_0^2 [c_1(1+z)^{k_1}+(1-c_1)(1+z)^{k_2}]\,, \qquad k_{1,2}=\frac{ 3\pm \sqrt{1+8j}}{2}\,,
\eeq
with $c_1$, for now, an arbitrary mathematical constant of integration to be determined through appropriate boundary conditions. By computing the deceleration function
\beq
\label{qconstj}
q(z) = -1+ \frac{H_0^2 [c_1 k_1(1+z)^{k_1}   +(1-c_1)k_2 (1+z)^{k_2}]}{H^2(z)}=-1+\frac{k_1H^2(z)+(1-c_1)\left( k_2 - k_1\right) H_0^2(1+z)^{k_2}}{2 H^2(z)}\,,
\eeq
we can actually relate this mathematical integration constant to the present day value of the deceleration parameter and the jerk parameter
\beq
\label{c1constj}
c_1 = 1+\frac{k_1-2(1+q_0)}{k_2-k_1}\,,
\eeq
where $k_{1,2}$ have been given in Eq.(\ref{generalconstj}). From the above equation, it is seen that the constant $c_1$ is completely kinematical in nature. It is determined in general by the present day values of the deceleration and jerk parameters, both of which can be estimated from astrophysical datasets without referring to any particular model.  

To reconstruct a modified gravity theory in the context of the late-time universe, one can proceed two ways. Firstly, one can reconstruct starting from a particular cosmic evolution $a(t)$ or $H(t)$ \cite{Dunsby:2010wg,He:2012rf,Ortiz-Banos:2021jgg,Choudhury:2019zod,Gadbail:2022jco,Gadbail:2023mvu}. One can also start by specifying a cosmographic condition \cite{Carloni:2010ph}, since a particular cosmic evolution can also be specified by a particular cosmographic condition \cite{Dunajski:2008tg} (e.g. the condition $j(z)=1$ corresponds to a $\Lambda$CDM-like evolution). On the contrary, one can also start the reconstruction starting from a given form of the dark energy equation of state parameter (e.g. $w_{\rm DE}(z)=-1$). The later is the so-called designer approach \cite{Song:2006ej,Pogosian:2007sw,Fay:2007uy,Kumar:2016bzd,Atayde:2021pgb,Albuquerque:2022eac}. The approach adopted in this paper is the first one. In particular, we make use of the cosmographic condition $j(z)=1$. We prefer the former approach because it is more grounded on available astrophysical datasets e.g. the luminosity distance-redshift data, which gives us bounds on the cosmographic parameters. 

The  Hubble rate (\ref{generalconstj}) smoothly reduces to the following $\Lambda$CDM-like behavior in the limit $\varepsilon \to 0$:
\be\label{HLCDM}
H^2(z) = H_0^2 h^2(z) = H_0^2 [c_1 (1+z)^3 + (1-c_1)]\,,
\ee
with $0<c_1<1$. We define a $\Lambda$CDM-like evolution to be a cosmic evolution which admits well-defined asymptotic past and asymptotic future epochs at which $H(z) \propto (1+z)^3$ and $H(z) \to const.$, respectively \cite{Chakraborty:2022evc}. It is important at this point to clarify the distinction between the $\Lambda$CDM-like evolution and the $\Lambda$CDM `model' itself. Of course, the $\Lambda$CDM model based on Einstein's GR does give rise to the $\Lambda$CDM-like evolution \eqref{HLCDM}, but it is possible for other models, in particular modified gravity based models to give rise to a $\Lambda$CDM-like evolution as well. See, for example, Ref.\cite{Chakraborty:2021jku} for $f(R)$ gravity or Ref.\cite{Gadbail:2022jco} for coincident gauge $f(Q)$ gravity reproducing $\Lambda$CDM-like evolution. How the mathematical integration constant $c_1$ is related to $\Omega_{m0}$ and $\Omega_{\rm{DE}0}$ is model dependent. For the particular $\Lambda$CDM `model', one has $c_1=\Omega_{m0}$, and hence the mathematical constant $c_1$ assumes the physical meaning of the present-day abundance of dark matter. However, in general, and one would obtain $c_1=c_1(\Omega_{m0},\Omega_{\rm{DE}0})$, and consequently its physical interpretation will change. The particular function $c_1(\Omega_{m0},\Omega_{\rm{DE}0})$, and consequently its physical interpretation, is, therefore, model dependent\footnote{For some discussion about this important conceptual point, the reader is referred to the discussion around Eq.(8) of Ref.\cite{Zhai:2013fxa} and \cite[Sec.2]{Chakraborty:2022evc}.}
While the parameter $c_1$ is a purely kinematical and model-independent parameter (Eq.\eqref{c1constj}), the available estimates of the present-day abundance of dark matter $\Omega_{m0}$ are model-dependent. The latter constitute possible sources to the Hubble tension \cite{Pedrotti:2024kpn}.  

It can be checked that for a $\Lambda$CDM-like evolution (\ref{HLCDM}) the deceleration parameter is
\be\label{qLCDM}
q = -1 - \frac{\dot{H}}{H^2} = \frac{1}{2} - \frac{3}{2}\left(\frac{1-c_1}{h^2}\right)\,,
\ee
where we have used the relation $\frac{d}{dt}\equiv-H(1+z)\frac{d}{dz}$. For $\Lambda$CDM-like evolution we have
\be 
\frac{dq}{Hdt} = (2q-1)(q+1)\,,
\ee
and consequently $q$ decreases monotonically from $q=\frac{1}{2}$ to $q=-1$. It can be calculated that $c_1$ is related to $q_0$ as
\be\label{c1-q0}
c_1 = \frac{2}{3}(1+q_0)\,,
\ee
whose relationship is consistent with (\ref{c1constj}) for $j=1$. Indeed, in the $\Lambda$CDM-like paradigm, {\it Cosmology is a search for two numbers}, $H_0$ and $q_0$, \cite{1970PhT....23b..34S}.  Therefore, we can express the $\Lambda$CDM-like cosmic evolution as
\be\label{HLCDMfinal}
H^2(z) = H_0^2 h^2(z) = H_0^2 \left[\frac{2}{3}(1+q_0)(1+z)^3 + \frac{1}{3}(1-2q_0)\right]\,,
\ee
such that it guarantees $H(z=0)=H_0,\,q(z=0)=q_0$.

Lastly, we propose a definition for the ``almost'' $\Lambda$CDM-like evolution. We define it such that all its cosmographic quantities and relations between them are ``almost'' like that in the $\Lambda$CDM-like evolution modulo regular terms with respect to a small parameter $\varepsilon$. 
Whether the same is also true for the reconstructed theory function $f(Q)$ or the dynamical connection function $\gamma(z)$ and the non-metricity variable $Q(z)$, is, however, not a trivial question due to the non-linearity of the field equations and demands explicitly investigation. In a clearer manner, the question can be posed as: does small deviation in cosmological kinematics also imply small deviation in cosmological dynamics?


\section{Basics of $f(Q)$ gravity theory}\label{sec:f(Q)}

In metric-affine theories, the affine connection can be decomposed into three distinct components as follows:

\begin{eqnarray}
\Gamma^\lambda_{\mu\nu}=\mathring{\Gamma}^\lambda_{\mu\nu}  +K^\lambda_{\mu\nu}+L^\lambda_{\mu\nu}\,,
\end{eqnarray}
where  $\mathring{\Gamma}^\lambda_{\mu\nu} $ is the Levi-Civita connection of the metric $g_{\mu\nu}$
           \begin{equation}
          \label{LeviCivita}
                 \mathring{\Gamma}^\lambda_{\mu\nu}  = \frac{1}{2} g^{\lambda \beta} \left( \partial_{\mu} g_{\beta\nu} + \partial_{\nu} g_{\beta\mu} - \partial_{\beta} g_{\mu\nu} \right) \,,
      \end{equation}
$ K^{\lambda}{}_{\mu\nu} $ is the contorsion tensor       
        \begin{equation}
          \label{Contortion}
                K^{\lambda}{}_{\mu\nu} =\frac{1}{2} g^{\lambda \beta} \left( -T_{\mu\beta\nu}-T_{\nu\beta\mu} +T_{\beta\mu\nu} \right)\, ,
      \end{equation}
and  $ L^{\lambda}{}_{\mu\nu} $ is the disformation tensor
     \begin{equation}
          \label{Disformation}
                L^{\lambda}{}_{\mu\nu} = \frac{1}{2} g^{\lambda \beta} \left( -Q_{\mu \beta\nu}-Q_{\nu \beta\mu}+Q_{\beta \mu \nu} \right)\,.
     \end{equation}
     
     It may be noted that the contorsion is constructed from the torsion tensor $T^{\lambda}{}_{\mu\nu}$ which is due to the anti-symmetric nature of the connection i.e.
      \begin{equation}
          \label{torsiontensor}
               T^{\lambda}{}_{\mu\nu}= \Gamma{}^{\lambda}{}_{\mu\nu}-\Gamma{}^{\lambda}{}_{\nu\mu}\,.
      \end{equation}
On the other hand, the disformation tensor is constructed from nonmetricity tensors as
\begin{equation}
          \label{nonmetricitytensor}
               Q_{\rho \mu \nu} \equiv \nabla_{\rho} g_{\mu\nu} = \partial_\rho g_{\mu\nu} - \Gamma^\beta_{\mu \rho}\,\, g_{\beta \nu} -  \Gamma^\beta_{\nu \rho } \,\,g_{\mu \beta}  \,.
      \end{equation}
      
   One may also defined  curvature tensor of the connection as
       \begin{eqnarray}
         \label{CurvatureTensor}
               R^{\sigma}\,_{\rho\mu\nu} &=& \partial_{\mu}\Gamma^{\sigma}\,_{\nu\rho}-\partial_{\nu}\Gamma^{\sigma}\,_{\mu\rho}+\Gamma^{\sigma}\,_{\mu\lambda}\Gamma^{\lambda}\,_{\nu\rho}-\Gamma^{\sigma}\,_{\mu\lambda}\Gamma^{\lambda}\,_{\nu\rho}\,.
         \end{eqnarray}
     
In analogy to the Ricci scalar obtained in GR due to the Levi-Civita connection,  in nonmetrcity theory, one can obtain the non-metricity scalar $Q$ due to the affine connection $\Gamma^\lambda_{\mu\nu}$ which is defined as the trace of the non-metricity tensor and given by \cite{BeltranJimenez:2017tkd}
\begin{align}\label{ns}
Q=Q_{\alpha\mu\nu}P^{\alpha\mu\nu}\,,
\end{align}
where $P_{~\mu\nu}^{\alpha}$ the non-metricity conjugate  given by
\begin{align}\label{P}
P_{~\mu\nu}^{\alpha}=-\frac{1}{2}L_{~\mu\nu}^{\alpha}+\frac{1}{4}\left(Q^{\alpha}-\overline{Q}^{\alpha}\right)g_{\mu\nu}-\frac{1}{4}\delta_{(\mu}^{\alpha}Q_{\nu)}\,,
\end{align}
with $Q_{\alpha}=Q_{\alpha\mu\nu}g^{\mu\nu}$, $\overline{Q}_{\alpha}=Q_{\mu\nu\alpha}g^{\mu\nu}$ as traces of the non-metricity tensor.

In this work, we consider an affine connection with vanishing curvature and torsion but a non-vanishing nonmetricity. Under such connection, a generic non-metricity  $f(Q)$ theory was formulated whose action is given by \cite{Hohmann:2019fvf,Zhao:2021zab,Hohmann:2021ast}
\begin{align}\label{action}
S=\int d^{4}x\sqrt{-g}\left[\frac{1}{2}f(Q)+\mathcal{L}_{m}\right]\,,
\end{align}
where $\mathcal{L}_{m}$ is the matter Lagrangian density, $g$ is the determinant of the metric $g_{\mu\nu}$, and $f(Q)$ is an arbitrary function of the invariant non-metricity scalar $Q$. The particular case $f(Q)=Q$ differs from Einstein-Hilbert action by only a boundary term. Therefore, at the level of the field equation, they are equivalent formulations \cite{BeltranJimenez:2019esp}. Hence, the theory given by $f(Q)=Q$ is called the Symmetric Teleparallel Equivalent of GR (henceforth STEGR in short). 


Since the action \eqref{action} is constructed from the metric and the connection, so there are two equations of motion. One of the equations of motion is due to variation with respect to the metric given by
\begin{align}\label{eom1}
f_{Q}G_{\mu\nu}+\frac{1}{2}g_{\mu\nu}(f_{Q}Q-f)+2f_{QQ}\nabla_{\alpha}QP_{~~\mu\nu}^{\alpha}=T_{\mu\nu}\,,
\end{align}
where $G_{\mu\nu}$ is the Einstein tensor and  $T_{\mu\nu}$ is the energy-momentum tensor of matter. Here, $f_{Q}=\partial f/\partial Q,f_{QQ}=\partial^{2}f/\partial Q^{2}$.

Another equation of motion due to variation with respect to the connection is
\begin{align}\label{eom2}
\nabla_{\mu}\nabla_{\nu}\left(\sqrt{-g}f_{Q}P_{~~~\sigma}^{\mu\nu}\right)=0\,.
\end{align}
Since, the matter Lagrangian is not coupled to the connection, the energy momentum tensor is separately conserved i.e. $\nabla_\mu T^{\mu\nu}=0$, here covariant derivative with respect to the Levi-Civita connection. On performing the Levi-Civita covariant derivative of (\ref{eom1}) and using Bianchi identity, one can obtain the connection equation (\ref{eom2}) \cite{Jarv:2018bgs}. Thus,
the two equations of motion are related to each other.

Let us mention an important physical viability condition here which will be used extensively later on to put constraints on the free parameters of the reconstructed $f(Q)$. The effective gravitational constant $\frac{1}{f_Q}$ dictates how the geometry responds to matter energy-momentum in $f(Q)$ gravity. The situation $f_Q<0$ may be interpreted as anti-gravity. The borderline case $f_Q=0$ implies infinite gravitational “constant” and corresponds to a singularity. Hence, physically viable models should obey the requirement $f_Q>0$ \cite{Guzman:2024cwa}.


\section{FLRW cosmology in $f(Q)$ theory}\label{sec:FLRW_f(Q)}

In this work, we consider the most general forms of the metric and connection in the context of symmetric telleparallel gravity,  
which are compatible with the cosmological principle on homogeneity and isotropy. The homogeneity and isotropy nature is determined by the invariance of the metric and the connection   with respect to the spatial rotational and translational transformations. Mathematically, the condition is equivalent to the vanishing Lie derivatives of the connection  with respect to the Killing vectors corresponding to spatial rotations and translation \cite{Hohmann:2021ast}. 

 A general form of spatially flat metric with homogeneous and isotropic condition is the FLRW metric given by
\begin{align}\label{FLRW}
ds^2 &= -dt^2 + a(t)^2 \left(dr^2 + r^2 d\theta^2 + r^2 \sin^2 \theta d\phi^{2} \right) \,.
\end{align}
The non-zero components of a  general non-vanishing torsion, affine connection  respecting homogeneity and isotropy are \cite{Hohmann:2021ast}
\begin{align*}
\Gamma_{\;tt}^{t} &=C_1,\quad\Gamma_{\;rr}%
^{t}=C_2,\quad\Gamma_{\;\theta\theta}^{t}=C_2\,r^{2},\quad
\Gamma_{\;\phi\phi}^{t}=C_2\, r^{2}\sin^{2}\theta\,,\\
\Gamma_{tr}^{r} &=C_3\,, \Gamma_{\theta\theta}^{r}  =-r~,~\Gamma_{\phi\phi}^{r}=-r\sin^{2}\theta~,\\  
\Gamma_{t \theta}^{\theta}  &  = C_3\,, \Gamma_{\theta r}^{\theta}=\frac{1}{r}\,,  \Gamma_{\phi \phi}^{\theta}=-\cos \theta \sin \theta\,,\\
\Gamma_{t \phi}^{\phi}  &  = C_3\,, \Gamma_{\phi r}^{\phi}=\frac{1}{r}\,,  \Gamma_{\theta \phi}^{\phi}=-\cot \theta \,,
\end{align*}
where $C_1, C_2, C_3$ are purely temporal functions. Imposing a curvature free condition, the above affine connection has  three possible branches depending on the values of $C_1, C_2, C_3$.

The first type of connection $\Gamma_1$ corresponds to:

\begin{align*}
(C_1,C_2,C_3)=(\gamma,0,0) \,.
\end{align*}

The second type of connection $\Gamma_2$ corresponds to
\begin{align*}
(C_1,C_2,C_3)=\left(\gamma+\frac{\dot{\gamma}}{\gamma},0,\gamma \right)\,,
\end{align*}

and the third type of connection $\Gamma_3$ corresponds to
\begin{align*}
(C_1,C_2,C_3)=\left(-\frac{\dot{\gamma}}{\gamma},\gamma,0 \right)\,.
\end{align*}

For details, the reader is referred to the paper \cite{Hohmann:2021ast}.

\section{Cosmological field equations}\label{sec:field_eqs}

To derive the cosmological field equations for the different branches of connections in $f(Q)$
gravity, we follow the conventions laid out in \cite{Paliathanasis:2023hqq}.
Specifically, we adopt the expression  $Q=-6H^2$ for $\Gamma_1$. In what follows, we write the field equations for each set of connections:
\begin{itemize}
    
\item  Branch 1 ($\Gamma_1$): \\

For the first branch, the nonmetricity scalar 
$Q$ is related to the Hubble parameter 
H as: 
\begin{align} 
Q &= -6H^2, \label{Q-1} 
\end{align}  
leading to the Friedmann and Raychaudhuri equations: 
\begin{subequations}
\begin{eqnarray}
    && 3H^{2}f_Q + \frac{1}{2}\left(f - Qf_Q\right) = \rho_{m}\,, \label{fried-1}
    \\
    && -2\frac{d}{dt}\left(H f_Q\right) - 3 H^{2} f_Q - \frac{1}{2}\left(f - Qf_Q\right) = 0\,. \label{raych-1}
\end{eqnarray}
\end{subequations}

\item Branch 2 ($\Gamma_2$): \\

In this branch, the scalar $Q$ acquires additional terms from the time-dependent function $\gamma(t)$: 
\begin{align} 
Q &= -6H^2 + 9\gamma H + 3\dot{\gamma}, \label{Q-2} 
\end{align}
modifying the Friedmann and Raychaudhuri equations as: 
\begin{subequations}
\begin{eqnarray}
    && 3H^{2}f_Q + \frac{1}{2}\left(f - Qf_Q\right) + \frac{3\gamma\dot{Q}f_{QQ}}{2} = \rho_{m}\,, \label{fried-2}
    \\
    && -2\frac{d}{dt}\left(H f_Q\right) - 3H^{2}f_Q - \frac{1}{2}\left(f - Qf_Q\right) + \frac{3\gamma\dot{Q}f_{QQ}}{2}=0\,, \label{raych-2}
\end{eqnarray}
\end{subequations}
with an additional equation governing the evolution of $Q$: 
\begin{align} 
    \dot{Q}^{2} f_{QQQ} + (\ddot{Q} + 3H \dot{Q}) f_{QQ} &= 0. \label{Qdot-2} 
\end{align} 

\item Branch 3 ($\Gamma_3$): \\

For the third branch, 
$Q$ depends on both the time-dependent function 
$\gamma(t)$ and the scale factor $a(t)$: 
\begin{align} 
Q &= -6H^2 + \frac{3\gamma}{a^2}H + \frac{3\dot{\gamma}}{a^2}, \label{Q-3} 
\end{align} 
with the corresponding Friedmann and Raychaudhuri equations: 
\begin{subequations}
    \begin{eqnarray}
    && 3H^{2}f_Q + \frac{1}{2}\left(f - Qf_Q\right) - \frac{3\gamma\dot{Q}f_{QQ}}{2a^{2}} = \rho_{m}\,, \label{fried-3}
    \\
    && -2\frac{d}{dt}\left(H f_Q\right) - 3H^{2}f_Q - \frac{1}{2}\left(f - Q f_Q\right) + \frac{\gamma\dot{Q}f_{QQ}}{2a^{2}} = 0\,, \label{raych-3}
\end{eqnarray}
\end{subequations} 
Here, the equation for $Q$ evolution is given by: 
\begin{align} 
\dot{Q}^{2} f_{QQQ} + \left[\ddot{Q} + \left(H + \frac{2\dot{\gamma}}{\gamma}\right)\dot{Q}\right]f_{QQ} &= 0. \label{Qdot-3} 
\end{align} 
    
On the redefinition $\gamma\rightarrow a^2 \gamma$ \footnote{Essentially what we do is to redefine $\gamma\to\tilde{\gamma}=\frac{\gamma}{a^2}$; since the dynamical connection function $\gamma(t)$ is a free function of time, we have this freedom of redefinition.}, the equations become
\begin{subequations}
    \begin{eqnarray}
    && Q = -6H^{2} + 9\gamma H + 3\dot{\gamma}\,, \label{Q-3_redef}
    \\
    && 3H^{2}f_Q + \frac{1}{2}\left(f - Qf_Q\right) - \frac{3\gamma\dot{Q}f_{QQ}}{2} = \rho_{m}\,, \label{fried-3_redef}
    \\
    && -2\frac{d}{dt}\left(H f_Q\right) - 3H^{2}f_Q - \frac{1}{2}\left(f - Q f_Q\right) + \frac{\gamma\dot{Q}f_{QQ}}{2} = 0\,, \label{raych-3_redef}
    \\
    && \dot{Q}^{2}f_{QQQ} + \left[\ddot{Q} + \left(5H + \frac{2\dot\gamma}{\gamma}\right)\dot{Q}\right]f_{QQ} = 0\,. \label{Qdot-3_redef}
    \end{eqnarray}
\end{subequations}

\end{itemize}

The energy conservation for pressureless matter reads as $\dot \rho =-3H \rho$ and it can be separately integrated into
\beq 
\label{eq1g1}
\rho_m=\rho_{m0} \left( \frac{a(t)}{a_0}\right)^3\rho_{m0}(1+z)^3\,.
\eeq
That the energy conservation equation remains the same as in GR is due to the fact that particles still follow geodesics governed by the Levi-Civita connection \cite[p.3]{BeltranJimenez:2019tme}.


\section{Reconstructing $\Lambda$CDM-mimicking $f(Q)$ with $\Gamma_1$}\label{sec:recon_Gamma1}

For the connection $\Gamma_1$, it is possible to formulate a general reconstruction method for $f(Q)$ gravity. By combining Eqs.\eqref{fried-1} and \eqref{raych-1}, we obtain  
\be\label{friedraych-1}
2 \frac{d}{dt}\left(H f_Q\right) = -\rho_m\,, 
\ee
which can be expressed in terms of the redshift $z$
\beq\label{eq}
\frac{d}{dz} (f_Q H)=\frac{\rho_{m0} (1+z)^2}{2H(z)}\,.
\eeq
Using the chain rule and  noting that for $\Gamma_1$, $Q=-6H^2$\, we can express $f_Q$ as 
\beq\label{eq1}
f_Q=-\frac{f'(z)}{12 H(z) H'(z)}\,,
\eeq
where a $^\prime$ denotes derivative with respect to $z$. If the background evolution $H(z)$ is known, Eq.\eqref{eq} can now be integrated to obtain $f_Q(z)$. With the relation \eqref{eq1}, we can now integrate Eq.\eqref{eq} to obtain
\beq
f'(z) =-18 H_0^2 \Omega_{m0} \,H'(z)\int \frac{(1+z)^2}{H(z)} dz +A_1 H'(z)\,,
\eeq
where $A_1$ is a constant of integration and we have used $\Omega_{m0}=\frac{\rho_{m0}}{3H_0^2}$. A second integration yields
\beq
f(z) =-18 H_0^2 \Omega_{m0} \int \left(H'(z)\int \frac{(1+z)^2}{H(z)} dz \right) dz +A_1 H(z)+A_2\,,
\eeq
where $A_2$ is another constant of integration.

As an example, let us now investigate the case of a $\Lambda$CDM-like evolution \eqref{HLCDM}. Then we have
\be
f(z) = -\frac{12 \Omega_{m0}}{c_1} \int H'(z) H(z)  dz +A_1 H(z)+A_2 = -\frac{6 \Omega_{m0}}{c_1} H^2(z)+A_1 H(z) +A_2\,.
\ee
Redefining $\frac{A_1}{\sqrt{6}}\rightarrow\beta$, $A_2\rightarrow-2\Lambda$, the reconstructed $f(Q)$ takes the form
\begin{equation}\label{recon-Gamma1}
    f(Q)= -2\Lambda+\frac{\Omega_{m0}}{c_1} Q+\beta \sqrt{-Q}\,.
\end{equation}
Note that the reconstruction method does not impose any constraints on the sign of the coefficient  $\beta$ or the the cosmological constant term $\Lambda$.


\subsection{Identifying the free parameters in the reconstructed $f(Q)$ ($\Gamma_1$)}\label{subsec:freeparams_Gamma1}

The way the reconstructed $f(Q)$ is expressed in Eq.\eqref{recon-Gamma1} may give one an impression that the reconstructed $f(Q)$ is characterized by four independent parameters $\Omega_{m0},c_1,\beta,\Lambda$. However, this is not the case. In this subsection we identify the free parameters of the reconstructed $f(Q)$ \eqref{recon-Gamma1}.

For the reconstructed theory \eqref{recon-Gamma1}, writing the Friedmann equation \eqref{fried-1} at $z=0$
\be\label{coeffs_rel_Gamma1}
\frac{3H_0^2 \Omega_{m0}}{c_1} = \rho_{m0} + \Lambda\,, \quad \Rightarrow \quad \frac{\Omega_{m0}}{c_1} = \Omega_{m0} + \frac{\Lambda}{3H_0^2}\,, \quad \Rightarrow \quad \Omega_{m0} = \frac{2\Lambda}{3H_0^2}\left(\frac{1+q_0}{1-2q_0}\right)\,,
\ee
\footnote{Even though the Friedmann equation at $z=0$ \eqref{coeffs_rel_Gamma1} looks very close to that in GR, in general it is different from GR. This is because in general one cannot naively set $\Omega_{m0}=c_1$, as neither of them are the free parameters as we have shown in this section.}
where at the last step we used the relation \eqref{c1-q0}. One can get the same equation from the Raychaudhuri equation \eqref{raych-1} as well, which is expected, since in the presence of a nondynamical connection like $\Gamma_1$, the Friedmann and the Raychaudhuri equations are not independent, given the conservation of the energy-momentum tensor. For a particular cosmological trajectory specified by a given value of $c_1$ (which delivers the observed value $q_0$), the above equation determines the value of the matter abundance parameter $\Omega_{m0}$ in terms of the model parameter $\Lambda$. In other words, the model parameter $\Lambda$, though not affecting the effective gravitational constant $\kappa_{\rm eff}=\frac{1}{f_Q}$, actually determines the value of the present day matter abundance $\Omega_{m0}$. Alternatively, one can demand that the class of reconstructed $f(Q)$ \eqref{recon-Gamma1} to deliver any preferred value of the present day matter abundance $\Omega_{m0}$ by suitably choosing the value of the model parameter $\Lambda$.

Therefore the reconstructed $f(Q)$ \eqref{recon-Gamma1} is characterized by only two free model parameters, namely $\beta$ and $\Lambda$, which are related respectively to the effective gravitational coupling and the present day value of the matter abundance. Taking into account the relation \eqref{c1-q0}, the reconstructed $f(Q)$ that reproduces the $\Lambda$CDM-like cosmic evolution \eqref{HLCDMfinal} can ultimately be expressed as
\begin{equation}\label{reconfinal-Gamma1}
    f(Q)= - 2\Lambda + \left(\frac{\Lambda/H_0^2}{1 - 2q_0}\right) Q + \beta \sqrt{-Q}\,.
\end{equation}

For the theory \eqref{reconfinal-Gamma1}, the Friedmann equation is
\be
3H^2 \left(\frac{\Lambda/H_0^2}{1 - 2q_0}\right) = \rho_{m0}(1+z)^3 + \Lambda\,.
\ee
If we demand that at high redshift, GR acts as a cosmological past attractor for the theory, then the relation $3H^2 \approx \rho_m$ must hold at high redshift, which demands setting the value of the model parameter $\Lambda$ as $\Lambda=H_0^2(1-2q_0)$. In that case we have the following identification
\be\label{iden-Gamma1}
\{\Omega_{m0},\Omega_{\Lambda0}\}=\left\lbrace\frac{2}{3}(1+q_0),\frac{1}{3}(1-2q_0)\right\rbrace\,,
\ee 
where we have defined $\Omega_{\Lambda0}\equiv\frac{\Lambda}{3H_0^2}$. For this particular case, the model parameter $\Lambda$ as well as the present day matter abundance $\Omega_{m0}$ is completely determined in terms of a single quantity, namely the cosmographic parameter $q_0$, whose value can be estimated in a model-independent manner from cosmographic data sets. $\beta$ is the only free parameter in the reconstructed theory that could not be determined in terms of $q_0$. 

Note that, setting $\Lambda=H_0^2(1-2q_0)$ is not necessary a-priori as long as obtaining a particular $\Lambda$CDM-like solution giving rise to the required value of $q_0$ (see \eqref{HLCDMfinal}) is concerned. The requirement $\Lambda=H_0^2(1-2q_0)$ comes from demanding that GR acts as a past cosmological attractor. If we do, however, set $\Lambda=H_0^2(1-2q_0)$, then the reconstructed $f(Q)$ \eqref{reconfinal-Gamma1} simplifies to
\begin{equation}\label{reconsimple-Gamma1}
    f(Q)= - 2H_0^2(1-2q_0) + Q + \beta \sqrt{-Q}\,.
\end{equation}
The above $f(Q)$ admits the $\Lambda$CDM solution \eqref{HLCDMfinal} with the present day value of the deceleration parameter $q_0$, as well as admits GR (more precisely, STEGR) as the cosmological past attractor.

One might wonder if successive derivatives of the Raychaudhuri equation might give rise to a new constraint that help us determine $\beta$. But this is not the case. Since Raychaudhuri equation is identically satisfied by the reconstructed $f(Q)$ \eqref{reconfinal-Gamma1} for all $z$, all it's successive derivatives are satisfied as well. $\beta$ is truly a free parameter.


\subsection{Constraints on the free model parameters of the reconstructed $f(Q)$ ($\Gamma_1$)}

Now that we have identified the free model parameters in the reconstructed $f(Q)$ \eqref{reconfinal-Gamma1}, namely $\beta$ and $\Lambda$, in this subsection we try to find possible constraints they need to satisfy. 
\begin{itemize}
    \item Eq.\eqref{coeffs_rel_Gamma1} pins down the sign of the model parameter $\Lambda$. Since $q_0\approx-0.55$ and $0<\Omega_{m0}<1$,
    \be
    0<\frac{\Lambda}{H_0^2}<\frac{3}{2}\left(\frac{1-2q_0}{1+q_0}\right)\,.
    \ee  
    \item One must have $f_Q>0$, otherwise the effective gravitational constant would be negative \cite{Guzman:2024cwa}. Substituting the explicit form of $f(Q)$ gives the condition
    \be\label{physviab_cond_Gamma1}
    \frac{\Omega_{m0}}{c_1} > \frac{\beta}{2\sqrt{6}H}\,, \qquad \text{or} \qquad \frac{\beta}{H_0} < 2\sqrt{6}\,\frac{\Lambda/H_0^2}{1-2q_0}\, h(z)\,\,\,\forall\, z \,.
    \ee
    where in the last step we have used Eqs.\eqref{coeffs_rel_Gamma1} and \eqref{c1-q0}. This, however, does not pin down the sign of $\beta$.
    \item One has the following two conditions coming from the Friedmann and the Raychaudhuri equation \eqref{fried-1} and \eqref{raych-1}
\be
f-2Qf_Q=2\rho_m>0\,, \qquad f_Q + 2Q f_{QQ} = \frac{\rho_m}{2H^2 (1+q)}>0\,.
\ee 
However, taking into account \eqref{c1-q0} and \eqref{coeffs_rel_Gamma1}, one can check that these equations are satisfied. Therefore, they do not put any additional constraint on the free model parameters.
\end{itemize}


\subsection{Effective gravitational coupling in the reconstructed $f(Q)$ ($\Gamma_1$)}

\begin{figure}[h!]
\centering
\includegraphics[width=8cm,height=6cm]{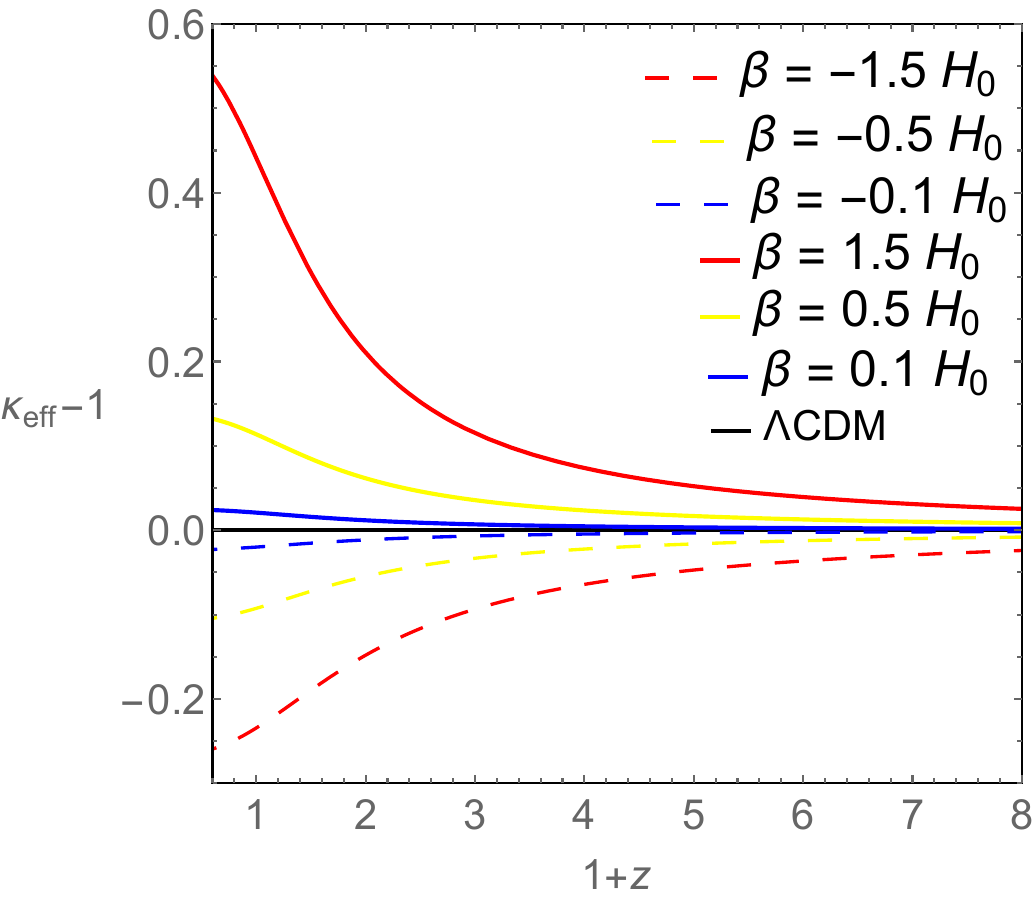}
\caption{Evolution of $\kappa_{\rm eff}$ for the 1-parameter family of reconstructed $f(Q)$ \eqref{reconsimple-Gamma1} that reproduces a $\Lambda$CDM-like cosmic evolution \eqref{HLCDMfinal} for the connection $\Gamma_1$. The evolution shows (STE)GR as a cosmological past attractor.}
\label{fig:kappa_eff_1}
\end{figure}

Lets us consider the $f(Q)$ in Eq.\eqref{reconsimple-Gamma1}. The effective gravitational constant is $\kappa_{\rm eff}=\frac{1}{f_Q}$ \footnote{$\kappa_{\rm eff}=8\pi G_{\rm eff}=\frac{8\pi G}{f_Q}$. Since we are working in the geometrized unit system, $\kappa_{\rm eff}=\frac{1}{f_Q}$.} with
\be\label{eq:kappaeff_gamma1}
f_Q(z) = 1 - \frac{\beta}{2\sqrt{-Q}} = 1 - \frac{\beta/H_0}{2\sqrt{6}h(z)}\,.
\ee
Now, suppose we consider the $f(Q)$ theory that reproduces the $\Lambda$CDM-like cosmology with $q_0=-0.55$, which corresponds to $c_1=0.3$. Because we have chosen the $f(Q)$ form from Eq.\eqref{reconsimple-Gamma1}, this also implies from \eqref{iden-Gamma1} that $\{\Omega_{m0},\Omega_{\Lambda0}\}=\{0.3,0.7\}$. Amazingly, we have arrived at the same value of $\Omega_{m0}$ and $\Omega_{\Lambda0}$ even without the underlying theory being GR. To respect the physical viability condition \eqref{physviab_cond_Gamma1} for all values of $z$, one needs to enforce the following condition on the model parameter $\beta$
\be\label{beta_range}
\frac{\beta}{H_0} < 2\sqrt{6}\,h_{\rm min} = 2\sqrt{6}\times 0.7 \approx 3.43\,. 
\ee 
For various values of the model parameter $\beta$ (taken from \cite{Albuquerque:2022eac}), all of which satisfy the above mentioned condition, we show in Fig.\ref{fig:kappa_eff_1} the evolution of the effective gravitational constant, along with that for the actual General Relativistic $\Lambda$CDM model.
It is possible to constrain the quantity $\kappa_{\rm eff}-1$ at $z=0$ using astrophysical datasets, which is expected to put some constraints on the model parameter $\beta$. On the other hand, since $\beta$ characterizes a deviation from the STEGR, one may expect distinct signatures at the level of matter perturbation evolution even though at the background level they are indistinguishable from the standard General Relativistic $\Lambda$CDM cosmographically. Some discussions about the latter is provided in the appendix \ref{app:distinguish}.

On the other way round, a prescribed value of $\beta$ sets the redshift interval on which our reconstructed theory is astrophysically applicable.
For example, the allowed time variation of the effective gravitational coupling is constrained from various  measurements, relying on data from Hulse-Taylor binary pulsar, helio-seismological
data, Type Ia supernova observations, and  the pulsation of the white dwarf G117-B15A, to be \cite{Ray:2005ia}
\be\label{kappa_variation}
|\dot \kappa_{\rm eff}/\kappa_{\rm eff}|< \bar C \approx 4.10 \cdot 10^{-11}\text{yr}^{-1} \qquad \text{up to $z<3.5$}.
\ee
This constraint has already become handy when performing the model selection for a holographic dark energy fluid with varying gravitational constant in \cite{Jamil:2009sq}, and it is specifically relevant in our case because it can be noted from Fig. \ref{fig:kappa_eff_1} that the steepest evolution of the effective gravitational coupling occurs at low redshift. 

Eq.(\ref{eq:kappaeff_gamma1}), together with (\ref{HLCDMfinal}) and the relationship between time and redshift derivatives, allows us to obtain
\beq
\bigg| \frac{\dot \kappa_{\rm eff}}{\kappa_{\rm eff}}\bigg|=\bigg| \frac{\dot f_Q}{f_Q}\bigg|=\bigg | \frac{(3h^2(z)+2q_0-1)\beta H_0}{2(2 \sqrt{6} H_0h(z)-\beta)h(z)} \bigg | = \frac{(1+z)^3 (1+q_0)|\beta| H_0}{(2 \sqrt{6} H_0h(z)-\beta)h(z)}\,,
\eeq
where we could move the absolute value at the denominator thanks to \eqref{beta_range}. Thus, we obtain the (implicit) condition on the redshift interval
\beq
\left(|\tilde \beta| - \frac{4\sqrt{6}}{3}\tilde{\bar C} \right)(1+q_0)(1+z)^3 - \frac{2\sqrt{6}}{3}\tilde{\bar C}(1-2q_0) + \tilde{\beta}\tilde{\bar C}\sqrt{\frac{2}{3}(1+q_0)(1+z)^3 + \frac{1}{3}(1-2q_0)} < 0\,,
\eeq
where we have defined $\tilde \beta=\beta/H_0,\,\tilde{\bar C}=\bar{C}/H_0$. Considering the model-independent values $q_0\approx-0.55$, $H_0\approx.1 \cdot 10^{-18}$ s$^{-1}$ \cite{Capozziello:2019cav}, the value of $\bar C$ in \eqref{kappa_variation} and recalling that 1 yr = $365\times 24 \times 60 \times 60$ s, we get that 
\beq
Y \equiv 0.45 (|\tilde \beta| - 2.03) (1+z)^3 -2.12+0.62 \tilde \beta \sqrt{0.30(1+z)^3 + 0.70} < 0
\eeq
should hold. As previously noticed, it can be easily seen that this condition holds towards past epochs by virtue of the specified range of $\beta$ \eqref{beta_range}, because the first term is the leading one. Direct computations allow us to confirm that the above condition also holds at the present epoch $z=0$ as long as $\beta$ is in the specified range \eqref{beta_range}. Moreover, the quantity $Y$ is a mototonically decreasing function of $(1+z)$, as evident from the following expression \footnote{The algebraic sum of the first two terms in \eqref{expression} is negative for our chosen $\tilde \beta$ values. Since the third term is negative for $\tilde \beta<0$, it is evident that $\frac{dY}{d(1+z)^3}<0$ for $\tilde{\beta}<0$. For positive $\tilde \beta$, the most disfavoured case would be for $z=0$, {\it e.g.} smallest possible denominator in the third term. In this case, an easy computation allows to confirm our claim that indeed $\frac{dY}{d(1+z)^3}<0$ for the chosen range of $\beta$.} 
\beq\label{expression}
\frac{dY}{d(1+z)^3} = 0.45|\tilde \beta|-0.91 + \frac{0.093 \tilde \beta }{ \sqrt{0.30(1+z)^3 + 0.70} }<0\,.
\eeq
Thus, all our chosen values of $\beta$ in Fig. \ref{fig:kappa_eff_1} remain within the observational bound on the variation of the effective gravitational coupling for the entire range of $z$ between $0<z<\infty$.


\subsection{Stability of a solution in the reconstructed $f(Q)$ for $\Gamma_1$}\label{subsec:stab_Gamma1}

In general, a solution $H(t)$ is characterized by a phase trajectory in the phase space, whereas fixed points characterize the asymptotic states of the solutions that are interpreted as cosmological epochs. For example, in the $\Lambda$CDM phase space, $\Lambda$CDM solutions are trajectories, whereas matter and $\Lambda$-dominated phases are fixed points (see \cite[Fig.2]{Boehmer:2014vea}). In this section, following the idea of \cite{Barrow:1983rx,delaCruz-Dombriz:2011oii}, we analyze the stability of the $\Lambda$CDM cosmological solution as a whole against small homogeneous and isotropic perturbations. 

Eq.\eqref{raych-1} can be written as
\be
2\dot{H} f_Q - 24\dot{H}H^2 f_{QQ} + 6H^2 f_Q + \frac{1}{2}f = 0\,.
\ee
For a given $f(Q)$, this is a differential equation in $H(t)$. Consider any given solution of this equation $H(t)$, and take a homogeneous and isotropic perturbation around it: $H(t)\rightarrow H(t)+\delta H(t)$. Substitute it back to the above equation. Keeping terms of only linear order we get (keep in mind that $Q=-6H^2 \Rightarrow \delta Q = -12H\delta H$)
\be
\dot{\delta H} + \lambda(Q) \delta H = 0\,.
\ee
where 
\be
\lambda(Q) = 3H - 36H\dot{H}\frac{f_{QQ} - 4H^2 f_{QQQ}}{f_Q - 12H^2 f_{QQ}} = 3H + 36(1+q)H^3 \frac{f_{QQ} + \frac{2}{3}Q f_{QQQ}}{f_Q + 2Q f_{QQ}}\,.
\ee
$\lambda(Q)$ determines the stability of the given solution $H(t)$. For stability of this solution with respect to small homogeneous and isotropic perturbations, one needs $\lambda(Q)>0$.

Consider, now, the theory given by Eq.\eqref{reconfinal-Gamma1}. For this theory, we have
\be
\lambda(Q)=3H\,,
\ee
which is always positive for an expanding cosmology. This implies that within the theory \eqref{reconfinal-Gamma1}, \emph{any} expanding cosmological solution, $\Lambda$CDM-like solution included, is always stable\footnote{Even though we reconstructed the theory \eqref{reconfinal-Gamma1} starting from the $\Lambda$CDM-like evolution, the theory can in principle admit as a solution not only the $\Lambda$CDM-like evolution but many other different kinds of evolution.}. If we start with a particular trajectory and move to a neighboring trajectory in its vicinity by a small homogeneous and isotropic perturbation at any point during its course of evolution, the new trajectory will show the same qualitative behavior (notably, the same past and future attractor) as the original trajectory we started with.


\subsection{Robustness of the reconstructed $f(Q)$ against slight deviation from the condition $j=1$}\label{subsec:rob_Gamma1}

Building upon our discussion from Sect.\ref{sec:ss2}, we will now demonstrate explicitly the robustness of our reconstructed $f(Q)$ in the framework of $\Gamma_1$, by investigating the a slightly more general scenario $j=1+\varepsilon=const.$ for ``small'' $\varepsilon$. This will be done by checking explicitly that there exists a smooth matching between the non-metricity function $Q$ and the reconstructed $f(Q)$ for the ``almost $\Lambda$CDM'' case  to their exactly $\Lambda$CDM-mimicking counterpart\footnote{For the case of $\Gamma_1$, the dynamical connection function is always set to zero as a geometrical property while constructing the $f(Q)$ theory: this applies regardless of the cosmic history and consequently a variation on the astrophysical value of the jerk does not bear any consequence on it. For the case of $\Gamma_2$ which we will consider in the next section, the smooth matching of the dynamical connection function will also have to be taken into consideration, allowing for potential discrimination between the two scenarios.}. The smoothness of the non-metricity directly follows from that of the Hubble rate via (\ref{Q-1}),  and it reads as
\beq
Q=-6H^2 = -6H^2_{\Lambda\rm CDM}+  \frac{2 \varepsilon [(z^3+3z^2+3z+2)c_1 - 1]}{3} \ln(1+z)+O(\varepsilon^2)\,.
\eeq
Next, for assessing the Raychaudhuri equation (\ref{raych-1}), we recall the deceleration function (\ref{qconstj}) for the cosmic rate reconstructed from a general constant jerk, where we should also use (\ref{eq1g1}), and re-write the redshift by using the Friedmann equation (\ref{fried-1}) with the conservation of dark matter, that is
\beq
(1+z)=\left( \frac{\rho_m}{\rho_{m0}} \right)^{1/3}=\left( \frac{f -2Qf_Q}{2 \rho_{m0}}  \right)^{1/3}\,.
\eeq
Then, from the Raychaudhuri equation
\begin{eqnarray}
2 H^2 (1+q) f_Q -24(1+q)H^4 f_{QQ}   -\frac{f}{2}-Qf_Q=0\, \quad \Rightarrow \quad
\frac{Q}{3}(1+q)(f_Q +2Qf_{QQ} ) +\frac{f}{2}-Qf_Q=0\,,
\end{eqnarray}
we can obtain the following closed-form differential equation for the reconstruction of the theory $f(Q)$:
\beq
\left[\frac{k_1 Q}{6} + (1-c_1)(k_1-k_2)H_0^2 \left( \frac{f -2Qf_Q}{2 \rho_{m0}} \right)^{k_2/3} \right](f_Q +2Qf_{QQ} ) + \frac{f}{2} - Qf_Q = 0\,.
\eeq
This equation can be recast in terms of the deviation $\varepsilon$ from the exact $\Lambda$CDM-like jerk parameter 
\beq
\label{generalg1}
\left[\frac{(3+\sqrt{9+8\varepsilon}) Q}{12} + (1-c_1)\sqrt{9+8\varepsilon}H_0^2 \left( \frac{f -2Qf_Q}{2 \rho_{m0}} \right)^{\frac{1}{2} - \frac{\sqrt{9+8\varepsilon}}{6}} \right](f_Q +2Qf_{QQ} ) + \frac{f}{2}-Qf_Q=0\,.
\eeq
The non-linear nature of the above equation and its dependence on the value of the jerk parameter $j=1+\varepsilon$ should be appreciated. It allows for an explicit confirmation that it consistently reproduces the case $\varepsilon=0$ with
\beq
f(Q) = - 2\Lambda + \frac{\Omega_{m0}}{c_1} Q + \beta \sqrt{-Q}\,,
\eeq
which is basically what we obtained in Eq.\eqref{recon-Gamma1}. By plugging the latter into the reconstruction differential equation (\ref{generalg1}) which holds for a general constant $j$, we obtain
\beq
\sqrt{9+8\varepsilon}\,\Omega_{m0}\left(\frac{1-c_1}{c_1}\right) \left(\frac{h^2}{c_1} - \frac{\Omega_{\Lambda0}}{\Omega_{m0}}\right)^{\frac{1}{2}-\frac{\sqrt{9+8\varepsilon}}{6}} - \frac{3}{2}\frac{\Omega_{m0}}{c_1}h^2 \left( \frac{\sqrt{9+8\varepsilon}}{3} - 1\right) - 3\Omega_{\Lambda0} = 0\,.
\eeq
For $|\varepsilon|\ll1$, linearizing the left hand side of the above equation with respect to $\varepsilon$, we arrive at the following condition
\beq
\frac{2\Omega_{m0}}{3c_1}\left[ (1-c_1) \left(2- \ln\left\vert \frac{h^2}{c_1} - \frac{\Omega_{\Lambda0}}{\Omega_{m0}}\right\vert\right) - h^2\right]\varepsilon + O(\varepsilon^2) \approx 0\,.
\eeq
Taking into consideration the expression \eqref{HLCDM}, the relations \eqref{c1-q0} and \eqref{coeffs_rel_Gamma1}, the above condition can be expressed as
\be
\left[\frac{1}{3}(1-2q_0)(1-\ln(1+z)) - \frac{2}{3}(1+q_0)(1+z)^3\right]\varepsilon + O(\varepsilon^2) \approx 0\,.
\ee 
For the term linear in $\varepsilon$, the quantity inside the square bracket is regular for $-1<z<\infty$. Therefore the term as a whole remains small, of the order $\mathcal{O}(\varepsilon)$. Hence, the reconstruction solution \eqref{recon-Gamma1} for $j=1$ ``almost solves'' the reconstruction differential equation for a constant $j=1+\varepsilon$ for any redshift $-1<z<\infty$. Hence, we conclude that our reconstructed LCDM-like $f(Q)$ \eqref{recon-Gamma1} holds within small observational error in the astrophysical estimate of $j$, and that ``almost $\Lambda$CDM'' can be achieved by ``almost the same $f(Q)$''.

Note that, we could not have inferred the same if the quantity inside the square bracket was divergent at any redshift within the range $-1<z<\infty$. Also, one can consider more accurate variation in the astrophysical estimate of the jerk parameter e.g. $j(z)=1+\varepsilon(z)$ (e.g. see \cite{Mukherjee:2020ytg}) or $j(z)=1+\varepsilon q(z)$, which also characterize an ``almost $\Lambda$CDM'' evolution. Such deviations are not considered here.


\section{Reconstructing $\Lambda$CDM-mimicking $f(Q)$ with $\Gamma_2$}\label{sec:recon_Gamma2}

For $\Gamma_2$, we were not able to formulate a generic reconstruction method unlike the $\Gamma_1$ case. Fortunately, we could perform the analytical reconstruction for the $\Lambda$CDM-like background evolution. Eq.\eqref{Qdot-2} can be written as
\ba
0 &=&
\frac{d}{dt}\left(\dot{Q} f_{QQ}\right)+ 3H (\dot{Q}f_{QQ}) 
=
\frac{d}{dt}\left(\dot{f_Q}\right)+ 3H \dot{f_Q} 
\nonumber\\
&=&
\frac{d^2 f_Q}{dt^2} + 3H \frac{d f_Q}{dt} = 0\,.
\label{ddfp}
\ea
Eq.(\ref{ddfp}) can be integrated for any generic cosmic history as \cite[Eq.(17)]{Yang:2024tkw}
\be
\label{iii}
\frac{d f_Q}{dt} = C a^{-3}\,,\quad \frac{d f_Q}{dz} = - C \frac{(1+z)^2}{H}\,, \quad f_Q(z) = -C \int \frac{(1+z)^2 dz}{H(z)}+D\,,
\ee
where $C$ is a dimensionful integration constant and $D$ is a dimensionless integration constant. From the second equality above, we see that the sign of the integration constant $C$ determines how the effective gravitational coupling $\kappa_{\rm eff}=\frac{1}{f_Q}$ evolves with time in an expanding universe. A positive (negative) $C$ implies that the effective gravitational coupling decreases (increases) with time\footnote{$C=0$ gives a constant effective gravitational coupling, which can be reduced to GR with a constant rescaling of the metric.}.  

Now, we try to reconstruct the $\Lambda$CDM-mimicking $f(Q)$ for $\Gamma_2$. Therefore, in whatever follows in this section, $H$ actually refers to the $\Lambda$CDM-like $H(z)$ as given in \eqref{HLCDM}, unless stated otherwise explicitly\footnote{When there is a scope of confusion, we will specify the $\Lambda$CDM-like $H$ as $H_{\Lambda\rm CDM}$.}. Then we get 
\ba
\label{recon1-Gamma2}
f_Q(z) = D - \frac{2 c}{3 c_1} \sqrt{c_1 (z+1)^3 + (1-c_1)} 
= D - \frac{2 c}{3 c_1} h(z)\,,
\ea
where we have defined $c\equiv\frac{C}{H_0}$, which is now dimensionless. Taking the sum of (\ref{fried-2}) and (\ref{raych-2}):
\beq
3\gamma \dot Q f_{QQ} -\rho_m -2 \frac{d}{dt}(f_Q H)=0\,,
\eeq
that can be re-written as
\beq
3\gamma \dot f_Q - \rho_m - 2 \frac{d}{dt}(f_Q H)=0\,.
\eeq
Now use the conservation for dust and the result for the reconstruction from (\ref{recon1-Gamma2}):
\beq
\label{int1}
\left(\frac{4}{3}h - \frac{\gamma}{H_0}- \frac{D c_1}{c}\right)\frac{2 c}{c_1}\frac{\dot h}{H_0} - 3\Omega_{m0}(1+z)^3 = 0\,.
\eeq
For the time derivative of the Hubble function we use
\beq
\frac{\dot h}{H_0} = (-h)(1+z)\frac{dh/dz}{H_0} = - \frac{3c_1 (1+z)^3}{2}\,.
\eeq
Consequently, we reconstruct
\beq\label{recon2-Gamma2}
\qquad \frac{\gamma}{H_0} = \frac{4}{3}h + \frac{\Omega_{m0}}{c}- \frac{D c_1}{c}\,.
\eeq
Therefore
\be\label{recon3-Gamma2}
\frac{\dot \gamma}{H_0^2} = - 2 c_1 (1+z)^3 = 2(1 - c_1 - h^2)\,.
\ee
Inserting the above equations into Eq.\eqref{Qdot-2}, we get
\begin{eqnarray}\label{recon4-Gamma2}
\frac{Q}{H_0^2} &=& -6h^2 +9 \frac{\gamma}{H_0} h + 3\frac{\dot{\gamma}}{H_0^2}\nonumber\\
&=& -6h^2 + 9h \left(\frac{4}{3}h + \frac{\Omega_{m0}}{c}-D \frac{c_1}{c}\right) + 6(1 - c_1 - h^2)\nonumber\\
&=& 6(1-c_1) + \frac{9h}{c} \left(\Omega_{m0} - c_1 D\right)\,.
\end{eqnarray}
Thus, both the reconstructed $\Lambda$CDM-like dynamical connection function and non-metricity for $\Gamma_2$ are linear in the Hubble parameter and monotonically increasing with respect to the redshift. Having in mind the previously established negativity of the constant $c$, we also understand that they are both allowed to vanish at some cosmic epoch as well as possibly switch their sign. For the nonmetricity, this constitutes  a sharp difference than its behavior for $\Gamma_1$, which is negative all along the comic history by construction (\ref{Q-1}),  and may provide a viable astrophysical route for discriminating among these scenarios. Next,  using Eq.\eqref{recon1-Gamma2} we get
\be
\label{recon5-Gamma2}
\frac{df}{dQ} = D+\frac{2 c^2}{27 c_1 \left(\Omega_{m0} - c_1 D\right)}\left[6(1-c_1)-\frac{Q}{H_0^2}\right]\,,
\ee
which allows us to ultimately reconstruct the $\Lambda$CDM-mimicking $f(Q)$ theory for the connection $\Gamma_2$
\beq\label{recon-Gamma2}
f(Q) = -2\Lambda + \left(D+\frac{4}{9} \frac{c^2 (1-c_1)}{c_1(\Omega_{m0}-c_1 D)} \right)Q - \frac{c^2}{27 c_1 (\Omega_{m0}-c_1 D) H_0^2}Q^2 \,.
\eeq
Screening mechanisms are likely to kill the effects of the cosmological constant $\Lambda$ on galactic length scales, in which context therefore our reconstructed theory is of quadratic nature, constituting a realistic and healthy model \cite[Eq.(13)]{Wang:2024eai}.


\subsection{Identifying the free parameters in the reconstructed $f(Q)$ ($\Gamma_2$)}

The way the reconstructed $f(Q)$ is expressed in Eq.\eqref{recon-Gamma2} may give an impression that the reconstructed $f(Q)$ is characterized by five independent parameters $\Omega_{m0},c_1,c,D,\Lambda$ ($H_0$ is not a ``free'' parameter as it is obtained from data in a model-independent way). However this is not the case. In this subsection we identify the free parameters of the reconstructed $f(Q)$ \eqref{recon-Gamma2}.

Eqs.\eqref{fried-2} and \eqref{raych-2} can be combined such that $\gamma$ is eliminated. In the resulting equation, using the reconstructed form of various quantities from Eqs.\eqref{recon1-Gamma2}, \eqref{recon2-Gamma2}, \eqref{recon3-Gamma2}, \eqref{recon4-Gamma2}, \eqref{recon5-Gamma2}, as well as the $\Lambda$CDM-like evolution \eqref{HLCDM}, after some steps of straightforward manipulations, we arrive at the following expression, which is, surprisingly, independent of the redshift $z$:
\be\label{coeffs_rel1_Gamma2}
4 c^2 \left(1 - c_1\right)^2+3 c_1^2 D \left(- 3 \left(1 - c_1\right) D+2 \lambda \right) - 6 c_1 \lambda  \Omega _{m0} +9 \left(1 - c_1\right) \left(\Omega _{m0}\right)^2 = 0\,,
\ee
where $\lambda \equiv \Lambda / H_0^2$.
The same equation can be obtained by evaluating the Friedmann equation \eqref{fried-2} at $z=0$. 
The above equation can be solved for $\Omega_{m0}$
\be\label{coeffs_rel2_Gamma2}
\Omega_{m0} = \frac{2}{3}\lambda\left(\frac{1 + q_0}{1 - 2 q_0}\right) \pm \frac{2}{3}\sqrt{(1 + q_0)^2 \left(\frac{\lambda}{1-2q_0} - D\right)^2 -\frac{c^2}{3}(1 - 2q_0)}\,, 
\ee
where at the last step we have used the relation \eqref{c1-q0}. For a particular cosmological trajectory specified by a given value of $c_1$ (which delivers the observed value $q_0$), Eq.\eqref{coeffs_rel2_Gamma2} determines the value of the matter abundance parameter $\Omega_{m0}$ in terms of the model parameters $\Lambda,c,D$. The model parameter $\Lambda$, although not affecting the effective gravitational constant $\kappa_{\rm eff}=\frac{1}{f_Q}$, contributes towards determining the value of the present day matter abundance $\Omega_{m0}$ via Eq.\eqref{coeffs_rel2_Gamma2}. Alternatively, one can demand that the class of reconstructed $f(Q)$ \eqref{recon-Gamma2} to deliver any preferred value of the present day matter abundance $\Omega_{m0}$ by suitably choosing the values of the free model parameters $\Lambda,c,D$ that satisfy Eq.\eqref{coeffs_rel1_Gamma2}. Therefore the reconstructed $f(Q)$ \eqref{recon-Gamma2} is characterized by three free model parameters, namely $c,\,D$ and $\Lambda$. Taking into account the relation \eqref{c1-q0}, the reconstructed $f(Q)$ that reproduces the $\Lambda$CDM-like cosmic evolution \eqref{HLCDMfinal} can ultimately be expressed as
\begin{equation}\label{reconfinal-Gamma2}
    f(Q) = -2\Lambda + \left[D + \frac{1}{3}\left(\frac{1-2q_0}{1+q_0}\right)\left(\frac{2c^2}{3\Omega_{m0}-2(1+q_0)D}\right)\right]Q - \left[\frac{c^2}{6(1+q_0)(3\Omega_{m0}-2(1+q_0)D)H_0^2}\right]Q^2\,,
\end{equation}
with $\Omega_{m0}$ given by Eq.\eqref{coeffs_rel2_Gamma2}. This clearly shows that the reconstructed $f(Q)$ is a 3-parameter model; the free parameters being $\Lambda,c,D$. 

The parameter $D$ is actually related to the redshift $z_*$ at which the reconstructed $f(Q)$ coincides with STEGR. Supposing that $f_Q(z_*) = 1$ at some redshift $z_*$, we get from Eq.\eqref{recon1-Gamma2}
\be
D = 1 - \frac{2 |c|}{3 c_1} h_*\,, 
\label{dSol}
\ee
where $h_*=h(z_*)$. Therefore, instead of $D$, $z_*$ (or equivalently $h_*$) can also be used as a free parameter of the reconstructed $f(Q)$ model. 

One may wonder, and justifiably so, what is the meaning of the two ``branches'' of $\Omega_{m0}$ given by the `$\pm$' sign in Eq.(\ref{coeffs_rel2_Gamma2}). Since $\Lambda,c,D$ are free parameters, it is possible for both branches to produce a physically consistent value of $\Omega_{m0}\in (0,1)$ for some range of values of the parameters $\Lambda,c,D$. This particular question demands some explanation. Applying Eq.(\ref{dSol}) to Eq.(\ref{recon2-Gamma2}),
we get
\be
\frac{\gamma}{H_0}  = \frac{4 h}{3} + \frac{2 h_*}{3} + \frac{\Omega_{m0} - c_1}{c}\,.
\label{gamma2om}
\ee
Eq.(\ref{gamma2om}) shows that the different value of $\Omega_{m0}$ leads to different constant term in the expression for $\gamma / H_0$. Inserting $D$ from Eq.(\ref{dSol}) into Eq.(\ref{recon4-Gamma2}), we get 
\be
\frac{Q}{H_0^2} = 6(1 - c_1 - 6 h h_*) - \frac{9 h (\Omega_{m0} - c_1)}{|c|}
= 6(1 - c_1) - h \left(\frac{9 (\Omega_{m0} - c_1)}{|c|}+ 6 h_*\right)\,,
\ee
which shows that $Q$ always varies with $h$ linearly, but with different rates for different values of $\Omega_{m0}$. In other words, different value of $\Omega_{m0}$, can lead to the same kinematic $\Lambda$CDM evolution, while different quantitative evolution of $Q(z)$\footnote{Different values of $\Omega_{m0}$ implies different rate of evolution for $Q(z)$, which may lead to different values of $\Omega_m$ at high redshift.}


Since $\Lambda,c,D$ are free parameters of the reconstructed $f(Q)$ \eqref{reconfinal-Gamma2}, we have some freedom to choose them. Let us consider the parameter choice
\be\label{coeffschoice-Gamma2}
\lambda = 1-2q_0\,, \qquad c^2 = \frac{3 (1- D)^2 \left(1 + q_0\right)^2}{1 - 2 q_0}\,.
\ee
It can be verified by straightforward substitution in Eq.\eqref{coeffs_rel2_Gamma2} that the above parameter choice yields a unique value for $\Omega_{m0}$; $\Omega_{m0}=\frac{2}{3}(1+q_0) = c_1$. The particular choice of the model parameters in Eq.\eqref{coeffschoice-Gamma2} gives us back the same identification as in Eq.\eqref{iden-Gamma1}
\be\label{iden-Gamma2}
\{\Omega_{m0},\Omega_{\Lambda0}\}=\left\lbrace\frac{2}{3}(1+q_0),\frac{1}{3}(1-2q_0)\right\rbrace\,.
\ee
Moreover, it can be checked that the choice \eqref{coeffschoice-Gamma2} also make the coefficient of the linear term in the reconstructed $f(Q)$ \eqref{reconfinal-Gamma2} unity, so that the reconstructed $f(Q)$ is of the form $f(Q)=-2\Lambda+Q+F(Q)$ with $F(Q)\propto-Q^2$. Indeed, in many works found in the literature $f(Q)$ is indeed taken in the form $f(Q)=Q+F(Q)$ \cite{Atayde:2021pgb,Albuquerque:2022eac,Sahlu:2022bgy,Sahlu:2024pxk,Yang:2024tkw}, to make it explicit that at some limit the $f(Q)$ matches with the STEGR with the correct Newtonian gravitational constant \footnote{Again, we remind, we are using a geometrized unit system with $\kappa=8\pi G=1$}. For the parameter choice \eqref{coeffschoice-Gamma2}, the reconstructed $f(Q)$ \eqref{reconfinal-Gamma2} simplifies to
\begin{equation}\label{reconsimple-Gamma2}
f(Q) = - 2H_0^2(1-2q_0) + Q - \frac{1}{4H_0^2}\left(\frac{1-D}{1-2q_0}\right)Q^2 \,.
\end{equation} 
The above $f(Q)$ is a 1-parameter $f(Q)$ model, parametrized by the single model parameter $D$, that admits the $\Lambda$CDM solution \eqref{HLCDMfinal} with the present day value of the deceleration parameter $q_0$ for the connection branch $\Gamma_2$.


\subsection{Constraints on the model parameters of the reconstructed $f(Q)$ ($\Gamma_2$)}

Now that we have identified the free model parameters in the reconstructed $f(Q)$ \eqref{reconfinal-Gamma2}, namely $\Lambda,c,D$, in this subsection we find the range of values of the parameters for which $f_Q$ can be unity at some particular redshift $z = z_* > 0$ and $f_Q$ is always positive. The requirement that $f_Q(z_*)=1$ at some redshift $z_*>0$ is a ``desirable'' physical demand ensuring that the reconstructed theory coincides with STEGR at some point in the past, although not strictly necessary if just reconstructing a $\Lambda$CDM-like background evolution is concerned. Beyond the cutoff redshift $z_*$, there is no point of considering a modified gravity anymore. In other words, we say that the effect of gravity modification becomes apparent only after the cutoff redshift $z=z_*$. The condition $f_Q(z)>0$ ensures the positivity of effective gravitational coupling.

\begin{itemize}
\item Firstly, we note from Eq.\eqref{recon1-Gamma2}\footnote{To play with $f_Q$, we have made use of the simpler expression \eqref{recon1-Gamma2}: $f_Q=D-\frac{2c}{3c_1}h$. We could have also determined $f_Q$ from the reconstructed $f(Q)$ \eqref{reconfinal-Gamma2}, which gives $f_Q = D + \frac{4}{9}\frac{c^2 (1-c_1)}{c_1(\Omega_{m0} - c_1 D)} - \frac{2 c^2}{27 c_1 (\Omega_{m0} - c_1 D)} \frac{Q}{H_0^2}$. Using the expression of $Q$ along a $\Lambda$CDM trajectory \eqref{recon4-Gamma2}, we see that the two expressions for $f_Q$ are in fact equivalent.} that if $C>0$, then $f_Q$ definitely becomes negative at some redshift irrespective of the value of $D$. To ensure that $f_Q > 0$ for all redshift, we demand that $C<0$, so that Eq.(\ref{recon1-Gamma2}) can be written as
\be
f_Q = \frac{2 |c|}{3 c_1} h + D\,. 
\label{fq-test}
\ee
where we remind that $c\equiv C/H_0$. Since $h(z)$ is a monotonically increasing function of $z$ for a $\Lambda$CDM-like evolution (Eq.\eqref{HLCDMfinal}), $h_*=h(z_*)>h(z=0)=1$. This gives an upper bound on $D$
\be
D < 1 - \frac{2 |c|}{3 c_1}\,. 
\label{D-upper}
\ee
\item Inserting $D$ from Eq.(\ref{dSol}) into Eq.(\ref{fq-test}), we get
\be
f_Q = 1 +  \frac{2 |c|}{3 c_1} (h - h_*)\,.
\label{fq-rep}
\ee
It follows from Eq.(\ref{fq-rep}) that $f_Q = 0$ when
\be
h = h_* - \frac{3 c_1}{2 |c|}\,.
\label{h=h*}
\ee
Hence, $f_Q$ will never vanish for $h > 0$ if $h$ in the left--hand--side of the above equation is negative or zero, yielding the condition
\be
\frac{3 c_1}{2 |c|} \geq h_*\,.
\label{cc1_hs}
\ee
The above condition, in conjunction with Eq.\eqref{dSol}, implies
\be\label{D-lower}
D\geq 0\,,
\ee
which is a lower bound on $D$.
\item Combining \eqref{D-upper} and \eqref{D-lower}, we get the following bound on the model parameters
\be\label{c-D-bound}
c < 0 \,, \qquad 0 \leq D < 1 - \frac{2 |c|}{3 c_1} < 1\,.
\ee 
If one trades $D$ for $z_*$ to be the free parameter, then one can instead write from Eq.\eqref{cc1_hs} the following bounds on the model parameters $c,z_*$
\be\label{c-bound}
-\frac{3 c_1}{2 h_*} = -\frac{3 c_1}{2 h(z_*)} \leq c <0\,.
\ee
Note that, $f_Q(z)>0$ can be directly verified from the positivity of $D$, or by substituting \eqref{cc1_hs} into \eqref{fq-rep}. The bound \eqref{c-D-bound} or \eqref{c-bound} ensure both $f_Q(z)>0$ for all $z$ and coincidence with STEGR at some $z_*$ in the past.
\item As of now no bound has been given on the model parameter $\Lambda$. Note that the relation \eqref{coeffs_rel2_Gamma2} is of the form $\Omega_{m0}=\Omega_{m0}(\Lambda,c,D)$\footnote{$H_0$ and $q_0$ are not a model parameters as they are obtained from cosmographic datasets in a model-independent manner.}. The condition for a real valued $\Omega_{m0}$ puts the following constraint on $\Lambda$.
\be\label{Lambda-bound}
\left\vert \frac{\Lambda}{H_0^2(1-2q_0)} - D \right\vert \geq \frac{|c|}{1+q_0}\sqrt{\frac{1-2q_0}{3}} = \frac{2|c|}{3c_1}\sqrt{1-c_1}\,.
\ee 
The condition $0 < \Omega_{m0} < 1$ puts a further bound on $\Lambda$. However, its form is complicated and we deem in not necessary to write down the explicit inequality here. The point we make here is that, given a choice of values of the parameters $c$ and $D$ according to \eqref{c-D-bound}, the choice of the free parameter $\Lambda$ is not completely arbitrary. It is to be chosen such that one gets a real valued $\Omega_{m0}$ from Eq.\eqref{coeffs_rel2_Gamma2} and that $0<\Omega_{m0}<1$.
\item Let us see what the above bounds imply for the simpler 1-parameter reconstructed $f(Q)$ \eqref{reconsimple-Gamma2}. Firstly, for this choice one has $\Omega_{m0}=\frac{2}{3}(1+q_0)$ with $q_0\approx-0.55$; see Eq.\eqref{iden-Gamma2}. So, the condition $0<\Omega_{m0}<1$ is automatically satisfied. 

Substituting $D$ from Eq.(\ref{dSol}) into the expression for $c$ in Eq.\eqref{coeffschoice-Gamma2}
\be\label{violation}
1 = \frac{3}{1 - 2 q_0} h_*^2
= \frac{1}{1 - c_1} h_*^2\,.
\ee
Since $c_1 < 1$, $h_*=h(z_*) > 1$ for any $z_*>0$, meaning the right hand side of the above equation is always $>1$. In that case, clearly the above equality can never be satisfied. This suggests that the value of $c$ from Eq.\eqref{coeffschoice-Gamma2} is not compatible with the value of $D$ in Eq.(\ref{dSol}). Physically, it means that at no point in the past the reconstructed $f(Q)$ \eqref{reconsimple-Gamma2} coincides with STEGR. However, the above equation can be satisfied if $h_*=h(z_*) < 1$, which can be achieved when $z_* < 0$. In the latter case the reconstructed $f(Q)$ \eqref{reconsimple-Gamma2} may coincide with STEGR at some point in the future.
\end{itemize}


\subsection{Effective gravitational coupling in the reconstructed $f(Q)$ ($\Gamma_2$)}\label{subsec:kappaeff_Gamma2}

Let us first consider the 1-parameter reconstructed $f(Q)$ in Eq.\eqref{reconsimple-Gamma2}. The effective gravitational constant is $\kappa_{\rm eff}=\frac{1}{f_Q}$ with
\be
f_Q = 1 - \frac{1}{2}\left(\frac{1-D}{1-2q_0}\right)\frac{Q}{H_0^2}\,.
\ee
with $\frac{Q}{H_0^2}$ given from Eq.\eqref{recon4-Gamma2}. Taking into account the relations \eqref{c1-q0}, \eqref{coeffschoice-Gamma2} and \eqref{iden-Gamma2}, as well as the bounds in \eqref{c-D-bound}, we see that $f(Q)$ can be written as
\be\label{eq:kappaeff_gamma2}
f_Q(z) = D + \sqrt{\frac{3}{1-2q_0}}\,(1-D)h(z)\,.
\ee 
Since $0 \leq D<1$ \eqref{c-D-bound}, the reconstructed $f(Q)$ asymptote to STEGR at $z\rightarrow-1$ irrespective of whatever is the value of $D$, implying that (STE)GR acts not as a past attractor but as a future attractor. This conclusion is in line with what we observed at the last item of the previous subsection.

\begin{figure}[h!]
\centering
 	\subfigure[]{%
 		\includegraphics[width=7cm,height=6cm]{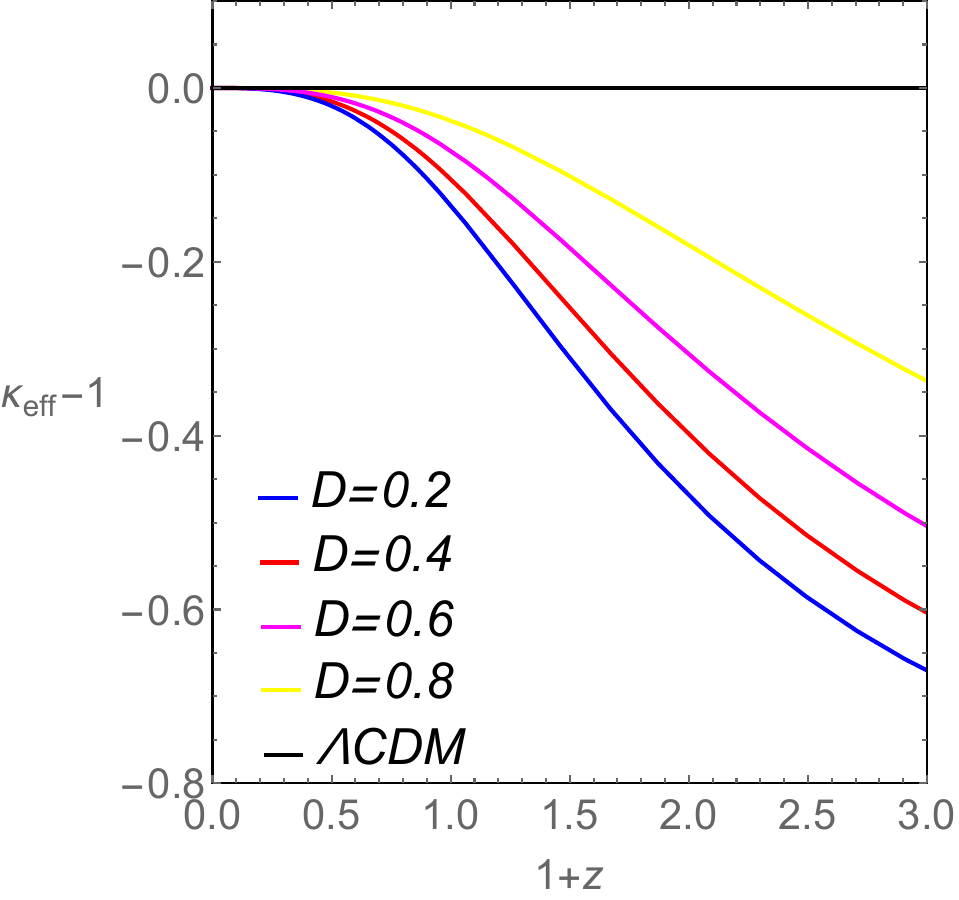}
 		\label{fig:kappa_eff_2a}}
 	\quad
 	\subfigure[]{%
 		\includegraphics[width=7cm,height=6cm]{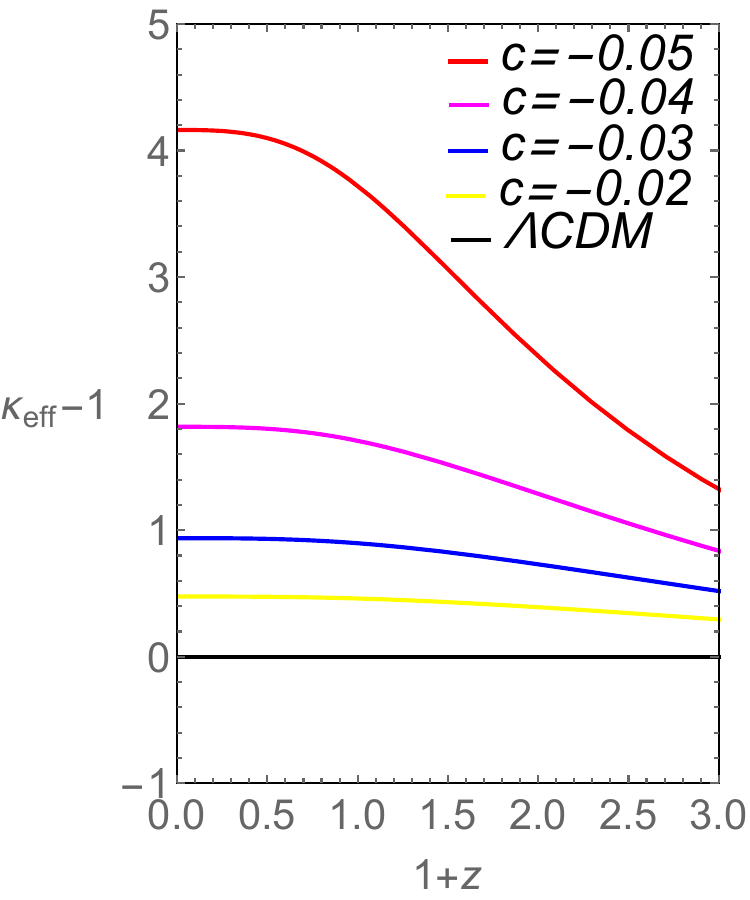}
 		\label{fig:kappa_eff_2b}}
\caption{(a) Evolution of $\kappa_{\rm eff}$ for the 1-parameter family of reconstructed $f(Q)$ \eqref{reconsimple-Gamma2} that reproduces a $\Lambda$CDM-like cosmic evolution \eqref{HLCDMfinal} for the connection $\Gamma_2$. The evolution shows STEGR as a cosmological future attractor. (b) Evolution of $\kappa_{\rm eff}$ for the reconstructed $f(Q)$ \eqref{reconfinal-Gamma2} that reproduces a $\Lambda$CDM-like cosmic evolution \eqref{HLCDMfinal} for the connection $\Gamma_2$. $z_*=5$ is chosen.}
\end{figure}

Suppose we consider one such $f(Q)$ theory that reproduces the $\Lambda$CDM-like cosmology with $q_0=-0.55$, which corresponds to $c_1=0.3$. Because we have chosen the $f(Q)$ form from Eq.\eqref{reconsimple-Gamma2}, this also implies from \eqref{iden-Gamma2} that $\{\Omega_{m0},\Omega_{\Lambda0}\}=\{0.3,0.7\}$. For four different values of $D$, the evolution of the effective gravitational constant is shown in Fig.\ref{fig:kappa_eff_2a}, along with that for the actual General Relativistic $\Lambda$CDM model.

We expect a matter dominated phase to be there at some high redshift. If the reconstructed $f(Q)$ does not reduce to STEGR around that redshift, there might be very different signatures coming from the analysis of matter perturbations. This may potentially render the connection branch $\Gamma_2$ ultimately non-viable, even though it reproduces a $\Lambda$CDM-like evolution with correct values of $q_0,\,\Omega_{m0}$.

Let us instead consider the generic reconstructed $f(Q)$ \eqref{reconfinal-Gamma2}, which is parametrized by three parameters $\Lambda,c,D$. In general $f_Q$ depends only on $c$ and $D$
\be 
f_Q(z) = D - \frac{2c}{3c_1}h(z)\,.
\ee 
If one trades $D$ for $z_*$ to be the free parameter, then one can write (\eqref{fq-rep})
\be 
f_Q(z) = 1 - \frac{c}{1+q_0}[h(z)-h(z_*)]\,.
\ee 
We can choose any value of $z_*$ that we want, but must choose a value of $c$ satisfying the bound in \eqref{c-bound} to ensure that the reconstructed $f(Q)$ coincides with STEGR at some high redshift, while also satisfying $f_Q(z) > 0$ for all $z$. For $z_*=5$ and different values of $c$, the evolution of the effective gravitational coupling is shown in Fig.\ref{fig:kappa_eff_2b}.


\section{Reconstructing $\Lambda$CDM-mimicking $f(Q)$ with $\Gamma_3$}\label{sec:recon_Gamma3}

To reconstruct the $\Lambda$CDM-mimicking $f(Q)$ for the case of $\Gamma_3$, we had to ultimately resort to numerical analysis. Eq.\eqref{Qdot-3_redef} can be written as
\be
\frac{d^2 f_Q}{dt^2} + \left(5H + 2\frac{\dot{\gamma}}{\gamma}\right) \frac{d f_Q}{dt} = 0\,,
\ee
which can be integrated to give
\be
\frac{d f_Q}{dt} = \frac{C}{a^{5}\gamma^2}\,,\quad \frac{d f_Q}{dz} = - \frac{C (1 + z)^{4}}{\gamma^2 H}\,.
\label{dfpdz-g3}
\ee
Unlike the case of $\Gamma_2$, for the case of $\Gamma_3$ the time-derivative of $f_Q$ depends on $\gamma$, so that it can be integrated only if the evolution of $\gamma$ is known. This brings in a lot of complications.
The evolution  equation for $\gamma$ can be obtain by firstly combining Eqs.\eqref{fried-3_redef} and \eqref{raych-3_redef} such that $f$ is eliminated:
\be
(\gamma +2 H) \dot{f}_Q + 2\dot{H}f_Q + \rho _m = 0\,.
\label{eq-g3}
\ee
Using Eq.\eqref{dfpdz-g3}, Eq.\eqref{eq-g3} can be written as
\be
\frac{C (1+z)^5 (\gamma +2 H)}{\gamma ^2} + 2 \dot{H} f_Q + \rho _m = 0\,.
\label{eq-mo}
\ee
The expression for $f_Q$ can be obtained by solving Eq.(\ref{eq-mo}) as
\be
f_Q = - \frac{C(1+z)^5 (\gamma + 2H) + \gamma^2 \rho _m}{2 \gamma^2 \dot{H}}\,.
\label{fp-sol}
\ee
Differentiating Eq.(\ref{eq-mo}) with respect to time, and then inserting the expression for $f_Q$ from Eq.\eqref{fp-sol} into the resulting equation,
we get
\be
(1 + z)^5 \gamma ^3 \left(\ddot{H}+3 H \dot{H}\right) \rho _m+\gamma  C \left(10 \dot{H} H^2+\dot{H} \left(\dot{\gamma }-4 \dot{H}\right)+2 H \ddot{H}\right)+\gamma ^2 C \left(\ddot{H}+5 H \dot{H}\right)+4 \dot{\gamma } C H \dot{H} = 0\,.
\label{eqDg}
\ee
We see that we have been able to obtain an evolution equation for $\gamma(z)$ that is decoupled from $Q(z)$. Since $\rho_{m}(z)=\rho_{m0}(1+z)^3$, the above equation can be used to solve for $\gamma(z)$ once an $H(z)$ is supplied. Once we know the solution $\gamma(z)$, Eqs.\eqref{Q-3_redef} and \eqref{fp-sol} can be used to solve for $Q(z)$ and $f_Q(z)$ respectively. One can, in principle, invert the expression of $Q(z)$ to solve for $z(Q)$, substitute it back into the expression of $f_Q(z)$ and get the reconstructed $f_Q(Q)=f_Q(z(Q))$, which can then be integrated to obtain $f(Q)$. Up to now, the relations we have derived are true for any generic cosmic history $H(z)$. Lets us now attempt the above exercise for a $\Lambda$CDM background. Therefore, in the rest of this section, unless stated otherwise explicitly, $H,\,q$ etc corresponds to that of a $\Lambda$CDM-like evolution as specified in Eq.\eqref{HLCDMfinal}\footnote{Again, if there is a scope of confusion, we will specify the dynamical quantities corresponding to the $\Lambda$CDM-like evolution with a subscript $\Lambda$CDM.}.
 
At this point we impose the condition that the Universe undergoes a $\Lambda$CDM-like evolution \eqref{HLCDM}, which corresponds to $j=1$. From the definition of the jerk parameter \eqref{j-def}, we see that $j=1$ implies $\ddot{H} = -3 H \dot{H}$. Substituting this, Eq.\eqref{eqDg} simplifies to
\be
\gamma  \left(\dot{\gamma }+4 H^2-4 \dot{H}\right)+2 \gamma ^2 H+4 \dot{\gamma } H = 0\,.
\label{eq4g0}
\ee
Surprisingly, explicit dependence on redshift and the matter density is washed away once we fix the background evolution to be $\Lambda$CDM-like. Also, for a $\Lambda$CDM-like evolution,
we can write
\be
\dot{H} = - \frac{3}{2} c_1 H_0^2 (1 + z)^3 = - \frac{3}{2} [H^2 - (1-c_1)H_0^2]\,,
\ee
so that Eq.(\ref{eq4g0}) becomes
\beq\label{eq4g}
\dot \gamma = -\frac{2H\gamma^2 + (10H^2 - 6(1-c_1) H_0^2) \gamma}{\gamma + 4H}\,.
\eeq
Substituting Eq.\eqref{eq4g} into Eq.\eqref{Q-3_redef}, we get
\be
Q = \frac{3 \left(6 \gamma(1-c_1)H_0^2 - 8 H^3 + \gamma^2 H\right)}{\gamma +4 H}\,.
\label{qq3}
\ee
Eq.\eqref{eq4g}, or equivalently Eq.\eqref{eq4g0} can be solved, at least numerically, for $\gamma(z)$ given the $\Lambda$CDM-like $H(z)$ from \eqref{HLCDMfinal}. Using the solution $\gamma(z)$ one can compute numerically $Q(z)$ and $f_Q(z)$ from Eqs.~(\ref{qq3}) and (\ref{fp-sol}). One can then, in principle, invert $Q=Q(z)$ to obtain $z=z(Q)$ and substitute it back to the expression of $f_Q(z)$ to obtain $f_Q(Q)=f_Q(z(Q))$. Integrating that will give the $\Lambda$CDM-mimicking $f(Q)$ for $\Gamma_3$.


\subsection{The system of equations for numerical reconstruction and the initial conditions}

For the purpose of numerical reconstruction, in proves to be useful to write the equations in terms of the following dimensionless quantities
\be 
x\equiv\frac{\gamma}{H}\,, \qquad \tilde{Q} \equiv \frac{Q}{H^2}\,, \qquad c\equiv\frac{C}{H_0^3}\,,\footnote{Readers are advised to take note of the difference in the definition of the dimensionless parameter $c$ between the cases $\Gamma_2$ and $\Gamma_3$.} \qquad N \equiv \ln a \,.
\ee 
In terms of the above variables, Eqs.~(\ref{fp-sol}), (\ref{eq4g0}) and (\ref{qq3}) can be written as
\ba
f_{Q} &=& \frac{c (x+2) (1+z)^5+3 h x^2 (1+z)^3 \Omega _{m0}}{3 h x^2 \left(c_1+h^2-1\right)}\,,
\label{fpx}
\\
\frac{d x}{d N} &=& \frac{x [(-1 + q) x-4]}{x+4}\,,
\label{dxdn}
\\
\tilde{Q} &=& \frac{3 \left[(2-4 q) x+x^2-8\right]}{x+4}\,.
\label{q3x}
\ea

Next, we need to specify the initial conditions, which we set by considering the reasonable assumption that the physics of gravity do not deviate much from GR during the matter dominated epoch, which occurs at high redshift; any modification of gravity becomes apparent at late time to drive cosmic acceleration. Then one can write
\be 
\Omega_{m0} (1 + z)^3 \simeq h^2(z) \simeq c_1 (1 + z)^3\,.
\ee 
In the above, the first equality comes from the fact that during the matter dominated epoch the physics of gravity is approximately GR so that the approximate equality $3H^2\approx\rho_m$ holds. The second equality comes from the fact that it occurs at high redshift. This gives rise to the identification $\Omega_{m0}=c_1$. The cosmic evolution during a matter dominated epoch in GR is characterized by the cosmographic condition $q = 1/2$, or equivalently $\dot{H}=- 3 H^2/2$. Let us assume that the reconstructed $f(Q)$ exactly coincides with STEGR at a redshift $z = z_*>0$ during this epoch, so that $f_Q(z_*)=1$.  Setting $f_Q = 1$, Eq.(\ref{fpx}) gives 
\be
x \simeq -2\,,\quad\mbox{at}\quad z \simeq z_*\,.
\label{g3-i}
\ee
Inserting $x = -2$ into Eq.(\ref{q3x}),
we get
\be 
\tilde{Q} \simeq - 6\,, \quad\mbox{at}\quad z \simeq z_*\,.
\label{q3-i}
\ee
Eqs.~(\ref{g3-i}) and (\ref{q3-i}) will be used to define the initial conditions during matter dominated epoch in the following analysis.


\subsection{Identifying the free parameters in the reconstructed $f(Q)$ ($\Gamma_3$)}\label{subsec:freeparams_Gamma3}

Up to this point, we already have two model parameters $\Omega_{m0}$ and $C$ in our analysis. Since we have to resort to numerical integration ultimately, we have to set the initial conditions at some redshift value $z_*$, which acts as another free model parameter\footnote{The situation is parallel to the case of $\Gamma_2$, where we have mentioned that $z_*$ can also be used instead of $D$ as the independent model parameter.}. Since our approach involves numerically finding $Q(z)$ and $f_Q(z)$, inverting $Q(z)$ to find $z(Q)$ and substituting it back to the expression of $f_Q(z)$ to obtain $f_Q(Q)$, we need to ultimately perform an integration to obtain $f(Q)$. This integration will produce the cosmological constant term $-2\Lambda$, which is another model parameter. So, there appears to be a total of four model parameters $\Lambda,c,z_*,\Omega_{m0}$. However, not all of them are independent. One can combine Eqs.\eqref{fried-3_redef} and \eqref{raych-3_redef} such that $\gamma$ is eliminated, yielding 
\be
(1+z)^{-5} \gamma ^2 \left[f_Q \left(6 H^2+6 \dot{H}-Q\right)+f+\rho _m\right]+6 C H = 0\,.
\label{f-eq}
\ee
In terms of the variable $x\equiv\frac{\gamma}{H}$, the above equation can be written as
\be 
f = -\frac{6 c (1+z)^5}{h x^2} + f_Q h^2 (6 q+\tilde{Q})-3 (1+z)^3 \Omega _{m0} \,.
\label{fx}
\ee
Eq.\eqref{fx}, evaluated at $z=0$, puts a constraint between the model parameters $\Lambda,c,z_*,\Omega_{m0}$, which can be used to determine $\Omega_{m0}$ in terms of of the other three. Therefore, ultimately, the reconstructed $f(Q)$ for the case of $\Gamma_3$ is in general parametrized by three model parameters just like in the case for $\Gamma_2$. The free model parameters are $\Lambda,c,z_*$.



\subsection{Approximate analytic reconstruction during the matter dominated epoch}\label{subsec:approx_Gamma3}

It is difficult to analytically solve the system of equations \eqref{fp-sol}, \eqref{dxdn}, \eqref{q3x} and obtain an analytic function $f(Q)$ valid for all $z$, precisely because of which we will go for a numerical reconstruction. However, before moving on to that, let us consider the solutions around the vicinity of the redshift $z=z_*$, while allowing for some deviations from GR (i.e. $c_1$ is not necessarily equal to $\Omega_{m0}$). It happens that we can reconstruct an approximate analytical $f(Q)$ during this epoch. The redshift $z=z_*$ is typically considered in a high enough redshift epoch so that one can safely approximate
\be\label{h_approx}
h^2 \approx c_1(1+z)^3 \,\,\Rightarrow\,\, q=\frac{1}{2}\,.
\ee
Setting $q = 1/2$, Eq.(\ref{dxdn}) becomes
\be
\frac{d x}{d N} = - \frac{x (x+8)}{2 x+8}\,.
\label{dxdn-mat}
\ee
This differential equation can be solved analytically, and the solutions are
\be
x(a) = -4 \pm \sqrt{16 + \frac{c_3}{a}}\,.
\label{xsolpm}
\ee
The constant of integration $c_3$ can be determined by setting $x(z_*)= -2$ according to Eq.(\ref{g3-i}). Hence, we get
\be
x(z) =  -4 + 2 \sqrt{4 - 3 \frac{1 + z}{1 + z_*}}\,,
\label{xsol}
\ee
where the plus  sign in Eq.(\ref{xsolpm}) is chosen to match with Eq.(\ref{g3-i}). It can be seen that this solution becomes imaginary for $z > \frac{1}{3}(1 + 4z_*)$. At $z = \frac{1}{3}(1 + 4z_*)$, we have $x = -4$ and consequently $d x / d N \to \infty$ according to Eq.(\ref{dxdn-mat}). For $q = 1/2$, Eq.(\ref{q3x}) becomes
\be
\tilde{Q} = \frac{3 \left(x^2-8\right)}{x+4}\,,
\label{q3-mat}
\ee
which shows that $Q \to \infty$ at $x = - 4$. This blowing up of $dx/dN$ and $Q$ is confirmed in numerical integration later on. Hence, the evolution of the system has a singularity during matter dominated epoch at the redshift $z = \frac{1}{3}(1 + 4z_*)$. Hence, an inherent assumption in our subsequent reconstruction attempt is that the matter dominated epoch starts after the redshift value $z < \frac{1}{3}(1 + 4z_*)$\footnote{In fact, this singularity can possibly be avoided in a more realistic scenario involving radiation, provided the transition from radiation to matter domination occurs at a redshift $z < \frac{1}{3}(1 + 4z_*)$.}.

According to Eq.(\ref{fpx}), it is possible for $f_Q$ to vanish for some value of $x$ and some values of the parameters $c$ and $\Omega_{m0}$. To find a range of parameter such that $f_Q$ never vanish within the interval $x > -4$, first we compute the value of $x$ that makes $f_Q$ vanish:
\be\label{def:y}
x_0 = - \frac 4y  \left(1 \pm  \sqrt{1 - y}\right)\,, \qquad y \equiv \frac{24 c_1 h}{c (1 + z)^2} \simeq \frac{24 c_1^{3/2}}{c \sqrt{1+z}}\,.
\ee
The last equality in the definition of $y$ comes from Eq.\eqref{h_approx}. To ensure that $f_Q$ never vanish for real $x$, one must demand $y > 1$, which places the following bounds on $c$
\be\label{c-bound_Gamma3} 
c < \frac{24 c_1^{3/2}}{\sqrt{1+z}} \,\,\, \forall z\in \left(0,\frac{1}{3}(1 + 4z_*)\right) \quad \Rightarrow c < \frac{12\sqrt{3} c_1^{3/2}}{\sqrt{1+z_*}}\,.
\ee  

To avoid a singularity of $x$, we analyze the solution near $z = z_*$. Expanding Eq.(\ref{xsol}) around $z = z_*$, we get
\be
x \simeq 1-\frac{3 (1+z)}{1+z_*}\,.
\label{x3sim}
\ee
Inserting Eq.(\ref{xsol}) into Eq.(\ref{q3-mat}) and expanding  the resulting expression around $z = z_*$, we get
\be
Q \simeq 3 H_*^2 \left(1-\frac{3 (1+z)}{1+z_*}\right)\,,
\label{qz*}
\ee
where $H_*=H(z_*)$. The expression for $f_Q$ can be obtained by substituting Eq.(\ref{xsol}) into Eq.(\ref{fpx}) yielding $f_Q$ around $z = z_*$ as
\be
f_Q \simeq 1 - \frac{C \left(z - z_*\right) \left(1+z_*\right)^4}{4 \left(H_*\right)^3}\,,
\label{fpz*}
\ee
Using Eqs.~(\ref{fpz*}) and (\ref{qz*}), we can write
\be
f_Q \simeq 1 + \frac{C \left(1+z_*\right)^5 \left(6 \left(H_*\right)^2+Q\right)}{36 \left(H_*\right)^5}\,,
\ee
which can be integrated as
\be\label{reconfinal_Gamma3_approx}
f(Q) \simeq -2\Lambda + Q + \frac{C Q \left(1+z_*\right)^5 \left(12 \left(H_*\right)^2+Q\right)}{72 \left(H_*\right)^5}\,, 
\ee
where $\Lambda$ is the constant of integration. Inserting the above equations for $x, Q$ and $f_Q$ into Eq.(\ref{f-eq}) and evaluating at $z=z_*$, we get
\be \label{f-eq_approx}
\left(C - \frac{2\Lambda H_*}{(1+z_*)^5}\right)\left[2(1+z_*)+\ln(1+z_*)\right] \simeq 0\,,
\ee
%
which can be satisfied when
\be
C = \frac{2 \Lambda  H_*}{\left(1+z_*\right)^5}\,,
\quad\ \Rightarrow \quad
\lambda \equiv \frac{\Lambda}{H_0^2} =  \frac{12 h_*}{y_*}\,, 
\ee
where $y_*=y(z_*)$. This relation suggests that $\Lambda$ relates to $\Omega_{m0}$ through $C$ and $z_*$. 

For the case of $\Gamma_2$, we had an integration constant $D$ arising in the expression of $f_Q$, which we could later relate to $z_*$. Also for the case of $\Gamma_3$ we can have a similar integration constant $D$. To show this, we insert $x$ from Eq.(\ref{x3sim}) into Eq.(\ref{dfpdz-g3}),
and perform an integration:
\be
f_Q = -\frac{C \left(1+z_*\right){}^3 \left(7 \left(1+z_*\right){}^2+2 (2 (1+z)-9) \left(1+z_*\right)+7\right)}{16 \left(H_*\right){}^3} + D\,,
\ee
where $D$ is the integration constant. Demanding that $f_Q = 1$ at $z = z_*$, 
we get
\be
D = 1 + \frac{C \left(11 \left(1+z_*\right)^2-18 \left(1+z_*\right)+7\right) \left(1+z_*\right)^3}{16 H_*^3}\,.
\ee

For the case of $\Gamma_2$, where we could carry out the analytic reconstruction for the $\Lambda$CDM-mimicking $f(Q)$, we could express $\gamma$ and $Q$ in terms of $H$ (Eqs.\eqref{recon2-Gamma2} and \eqref{recon4-Gamma2}). We can do the same here for the approximate reconstructed $f(Q)$ for $\Gamma_3$ in the vicinity of the matter-dominated epoch. For the sake of completeness we give the respective expressions below. Since the matter dominated epoch occurs at a high redshift, $h^2 \approx c_1(1+z)^3$ holds. From \eqref{xsol} and \eqref{q3-mat}, straightforward substitution yields
\be 
\frac{\gamma}{H_0} = -4h\left(1 - \sqrt{1 - \frac{3}{4}\left(\frac{h}{h_*}\right)^{2/3}}\right)\,, \qquad \frac{Q}{H_0^2} = 2h^2 \left[\left(1 - \frac{3}{4}\left(\frac{h}{h_*}\right)^{2/3}\right)^{-1/2} + 2\left(1 - \frac{3}{4}\left(\frac{h}{h_*}\right)^{2/3}\right)^{1/2} -4\right]\,.
\ee


\subsection{Numerical reconstruction}

\begin{figure}[H]
\centering
 	\subfigure[]{%
 		\includegraphics[width=7cm,height=6cm]{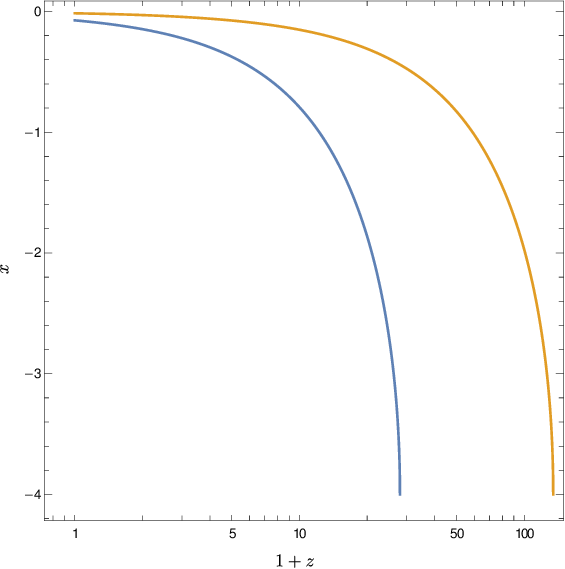}\label{fig:x3}}
 	\quad
 	\subfigure[]{%
 		\includegraphics[width=7cm,height=6cm]{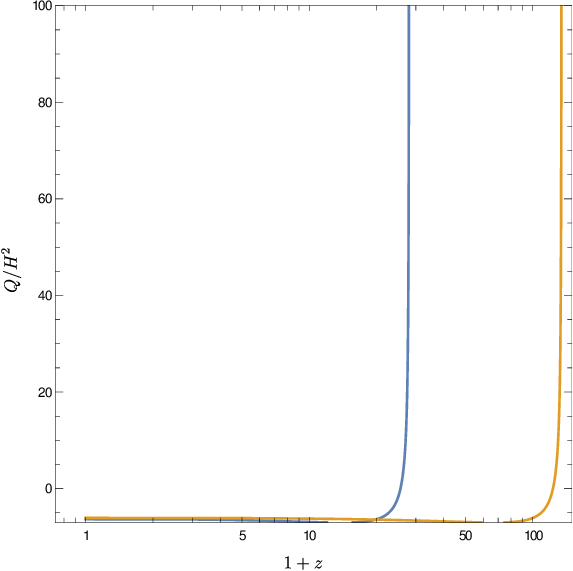}\label{fig:q3}}
\caption{Plots (a) and (b) shows the evolution of $x\equiv\frac{\gamma}{H}$ and $\tilde{Q}=\frac{Q}{H^2}$ as functions of $z$ as obtained by solving the equations \eqref{dxdn} and \eqref{q3x} for the parameter value $c=0.1$. The blue and the orange plots correspond to $z_*=20$ and $z_*=100$. The initial conditions are chosen according to Eqs.\eqref{g3-i} and \eqref{q3-i}.} 
\end{figure}

\begin{figure}[H]
\centering
 	\subfigure[]{%
 		\includegraphics[width=7cm,height=6cm]{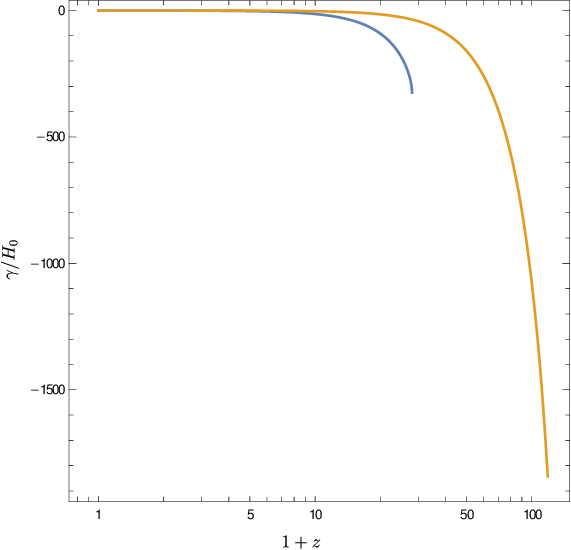}\label{fig:gam3H0}}
 	\quad
 	\subfigure[]{%
 		\includegraphics[width=7cm,height=6cm]{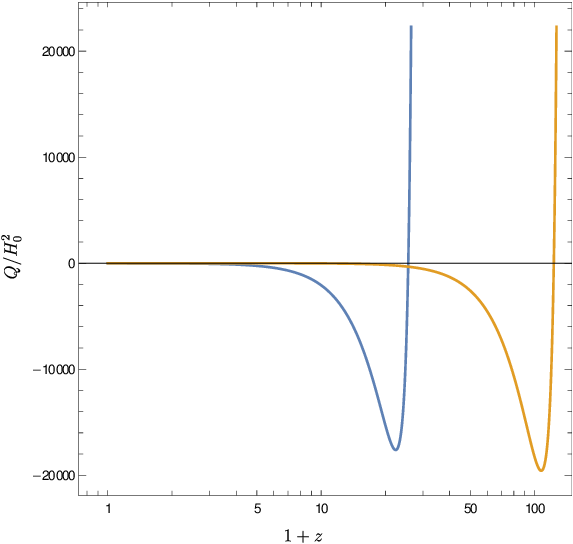}\label{fig:q3H0}}
\caption{Plots (a) and (b) shows the evolution of $\frac{\gamma}{H_0}$ and $\frac{Q}{H_0^2}$ corresponding to the plots of $x$ and $\tilde{Q}$ in Figs.\ref{fig:x3} and \ref{fig:q3}. For the plot of $\frac{Q}{H_0^2}$,  the value of $Q$ is divided  by $100$ (i.e. corresponding to $z_*=100$) for the orange plot to fit both of them in the same frame.} 
\end{figure}

\begin{figure}[H]
\centering
\subfigure[]{%
 		\includegraphics[width=7cm,height=6cm]{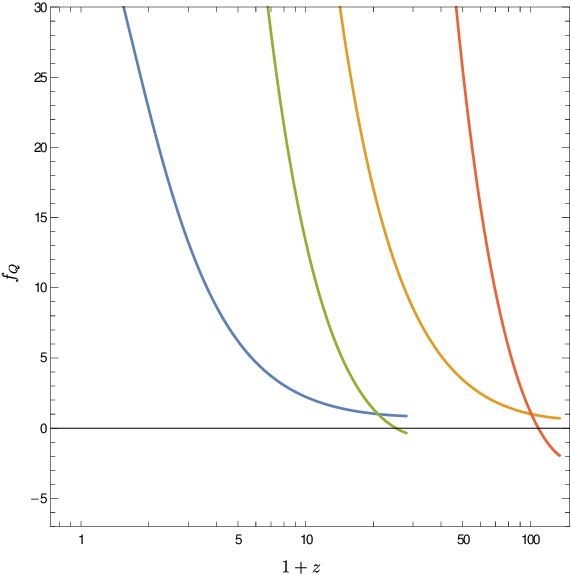}\label{fig:fp3}}
 	\quad
 	\subfigure[]{%
 		 		\includegraphics[width=7cm,height=6cm]{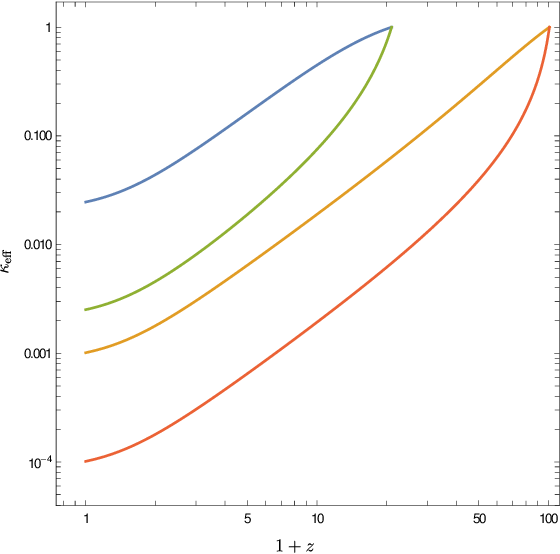}\label{fig:kp3}}
\caption{The plots for the $f_Q$ and $\kappa_{\rm eff}=\frac{1}{f_Q}$ as obtained from Eq.\eqref{fpx}. To show that the model parameter $c$ should be properly chosen, we have considered two values of $c$, $c=0.1$ and $c=1$ for each value of $z_*$. The blue and the green plots represent the cases $z_* = 20$. The orange and the red plots represent the cases $z_* = 100$. The blue and the orange plots represent the cases $c = 0.1$. The green and the red lines represent the cases $c = 1$. For the latter case we can see that $f_Q$ changes sign at some high redshift, corresponding to a discontinuity in $\kappa_{\rm eff}$.}
\label{fig:plots-g3}
\end{figure}

\begin{figure}[H]
 	\centering
 	 	\subfigure[]{%
 		\includegraphics[width=7cm,height=6cm]{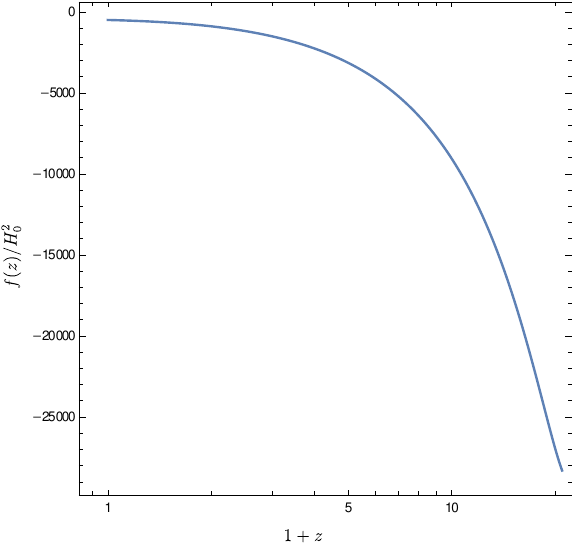}}
 		\quad
 	\subfigure[]{%
 		\includegraphics[width=7cm,height=6cm]{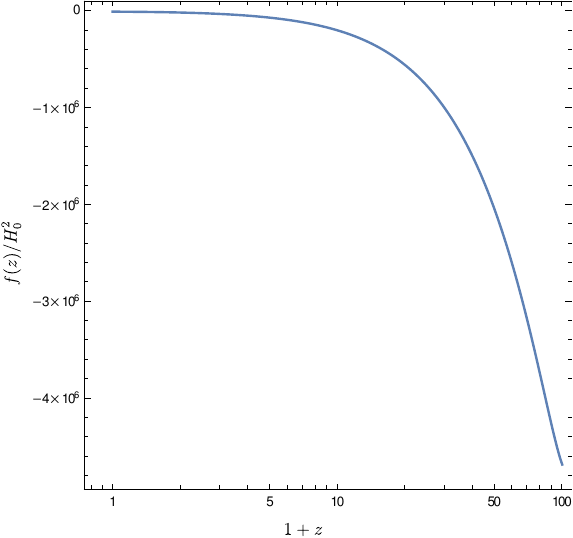}}
\caption{The plots of $f$ as a function of the redshift $z$ for $c = 0.1$ within the range $z\in \left(0,z_*\right)$. The left panel shows the case of $z_* = 20$, while the right panel shows the case of $z_* = 100$.} 
 	\label{fig:fz3}
 \end{figure}

\begin{figure}[H]
 	\centering
 	 	\subfigure[]{%
 		\includegraphics[width=7cm,height=6cm]{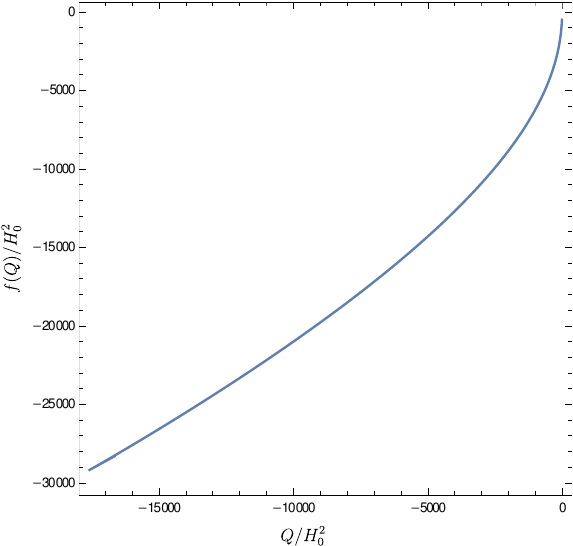}\label{fig:fq31}}
 		\quad
 	\subfigure[]{%
 		\includegraphics[width=7cm,height=6cm]{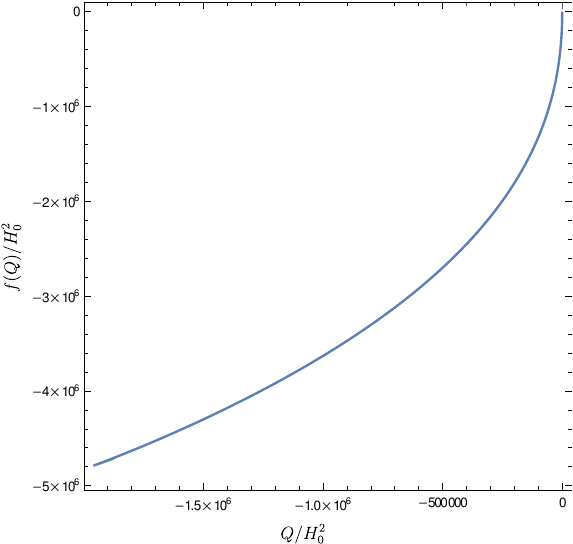}\label{fig:fq32}}
\caption{The plots of $f$ as a function of $Q$ for $c = 0.1$. The left panel shows the case of $z_* = 20$, while the right panel shows the case of $z_* = 100$.} 
 	\label{fig:fplot-g3}
 \end{figure}

Let us now do the full numerical reconstruction, setting $\Omega_{m0} = c_1 = 0.3$. We choose two values of $z_*$ for comparison: $z_*=20$ and $z_*=100$.
For the purpose of the numerical integration $z$ is limited within the range $\left(0,\frac{1}{3}(1+4z_*)\right)$ to avoid the singularity at $z = \frac{1}{3}(1+4z_*)$. The initial conditions are set according to Eqs.\eqref{g3-i} and \eqref{q3-i}. The plots of $x=\frac{\gamma}{H}$ and $\tilde{Q}=\frac{Q}{H^2}$ vs $z$ are shown in Figs.\ref{fig:x3} and \ref{fig:q3}.
In Figs.\ref{fig:gam3H0} and \ref{fig:q3H0} the corresponding $\frac{\gamma}{H_0}$ and $\frac{Q}{H_0^2}$ are plotted.
 It follows from the plots that $\tilde{Q}$ grows rapidly and changes sign when $x \to -4$. 
To show that $f_Q$ can vanish during cosmic evolution leading to a divergence of $\kappa_{\rm eff}$ if $c$ is not properly chosen, we numerically compute $f_Q$ for two choices of the parameter $c$: $c = 0.1$ and $c=1$. For our particular choice of the parameter values $\Omega_{m0}=c_1=0.3$ and for both the cases $z_*=20$ and $z_*=100$, it can be checked that $c = 0.1$ satisfies the bound in Eq.\eqref{c-bound_Gamma3}, while $c=1$ doesn't. Accordingly, the plots in Fig.\ref{fig:fp3} show that $f_Q$ can change sign from positive to be negative when $z > z_*$ for the case of $c = 1$. We show the plots for $\kappa_{\rm eff}$ from $z_*$ to the present epoch in Fig.\ref{fig:kp3}. 
Finally, the plots for $f$ as functions of the redshift and $Q$ are shown in Figs.\ref{fig:fz3} and \ref{fig:fplot-g3}, respectively. For these plots, we just consider the value $c=0.1$, which ensures the positivity of the effective gravitational coupling within the redshift range considered.

\section{Summary and discussions}\label{sec:sum_disc}

It is well known  that there are three symmetric teleparallel connection branches that respect homogeneity and isotropy \cite{Hohmann:2021ast}. Most cosmological studies in $f(Q)$ gravity have focused on the trivial connection branch $\Gamma_1$, where Cartesian coordinates serve as the coincidence gauge. Only in recent years have alternative connection branches garnered significant attention. In this work, we explore the possibility of reproducing an exact $\Lambda$CDM-like background evolution \eqref{HLCDMfinal} within the framework of $f(Q)$ gravity, considering each of the three connection branches individually. While this question has been previously addressed for the trivial branch $\Gamma_1$ in \cite{Gadbail:2022jco}—with which our results are consistent—we extend the analysis to the two nontrivial branches. Furthermore, we carefully identify the free parameters of the reconstructed $f(Q)$ for each case and examine the possible constraints among these parameters. The key takeaways from our reconstruction analysis are as follows:

\begin{itemize}
	
\item \textbf{For the trivial connection { \boldmath$\Gamma_1$}}, the general $f(Q)$ theory that matches $\Lambda$CDM evolution is given by the following 2-parameter family \eqref{reconfinal-Gamma1}:
\begin{equation}
    f(Q)= - 2\Lambda + \left(\frac{\Lambda/H_0^2}{1 - 2q_0}\right) Q + \beta \sqrt{-Q}\,,
\end{equation}
with $\beta$ and the cosmological constant term $\Lambda$ acting as the two free parameters. The physical viability condition, $f_Q > 0$, imposes the following  bounds on the free parameters:
\be 
0<\frac{\Lambda}{H_0^2}<\frac{3}{2}\left(\frac{1-2q_0}{1+q_0}\right)\,, \qquad \frac{\beta}{H_0} < 2\sqrt{6}\,\frac{\Lambda/H_0^2}{1-2q_0}\, h(z)\,\,\,\forall\, z \,.
\ee 
Demanding that STEGR acts as a cosmological past attractor simplifies the above to a 1-parameter family :
\begin{equation}
    f(Q)= - 2H_0^2(1-2q_0) + Q + \beta \sqrt{-Q}\,, \qquad \frac{\beta}{H_0} < 2\sqrt{6}h(z)\,\,\,\forall\, z \,.
\end{equation}

\item \textbf{ For the connection { \boldmath$\Gamma_2$}}, the general $f(Q)$ theory matching $\Lambda$CDM evolution is given by the following 3-parameter family \eqref{reconfinal-Gamma2}:
\begin{equation}
    f(Q) = -2\Lambda + \left[D + \frac{1}{3}\left(\frac{1-2q_0}{1+q_0}\right)\left(\frac{2c^2}{3\Omega_{m0}-2(1+q_0)D}\right)\right]Q - \left[\frac{c^2}{6(1+q_0)(3\Omega_{m0}-2(1+q_0)D)H_0^2}\right]Q^2\,,
\end{equation} 
with $c,\,D$ and the cosmological constant term $\Lambda$ acting as the three free parameters, and $\Omega_{m0}$ given in terms of the free parameters as \eqref{coeffs_rel2_Gamma2}. One can trade $D$ for $z_*$ as a free parameter (\eqref{dSol}), $z_*$ being the redshift at which the reconstructed $f(Q)$ coincides with STEGR. The physical viability requirement $f_Q>0$ for all $z$ and the physically desirable requirement that the reconstructed $f(Q)$ coincides with STEGR at some point in the past (i.e. $z_*>0$) is ensured by the following bounds on the free parameters :
\be 
-\frac{3 c_1}{2 h(z_*)} < c < 0 \,, \qquad 0 < D < 1 - \frac{2 |c|}{3 c_1} < 1\,.
\ee 

With suitable parameter choices \eqref{coeffschoice-Gamma2} it is possible to obtain a simpler 1-parameter family \eqref{reconsimple-Gamma2}:
\begin{equation}
f(Q) = - 2H_0^2(1-2q_0) + Q - \frac{1}{4H_0^2}\left(\frac{1-D}{1-2q_0}\right)Q^2 \,.
\end{equation}
However, this simpler family has the flaw that STEGR acts not as a cosmological past attractor, but rather as a cosmological future attractor ($f_Q(z\to-1)=1$), as shown in section \ref{subsec:kappaeff_Gamma2}. Such a theory will significantly deviate from STEGR during the matter dominated epoch in the past, leading to a signature in the structure formation scenario very different than observed. 

\item \textbf{ For the connection { \boldmath$\Gamma_3$}}, we were able to decouple the evolution equation for the dynamical connection for a generic cosmic history (Eq.\eqref{eqDg}) as well as for the $\Lambda$CDM-like evolution in particular (Eq.\eqref{eq4g}). This enabled us to perform a subsequent numerical reconstruction. The result of the numerical reconstruction is shown in Fig.\ref{fig:fplot-g3}. The reconstructed $f(Q)$ is a 3-parameter family in general, as discussed in section \ref{subsec:freeparams_Gamma3}, parametrized by $c,\,z_*$ and the cosmological constant term $\Lambda$. An approximate analytic form of the reconstructed $f(Q)$ during the matter dominated epoch is given by \eqref{reconfinal_Gamma3_approx}:
\be 
f(Q) \simeq -2\Lambda + Q + \frac{c Q \left(1+z_*\right)^5 \left(12 h_*^2 + Q/H_0^2\right)}{72 h_*^5}\,, \qquad h_*=h(z_*)\,.
\ee 

By appropriately choosing the free parameters,  it is possible to ensure the positivity of $f_Q$ within the range $z \in (0,z_*)$ (see Fig.\ref{fig:fp3}).
\end{itemize}

To alleviate the cosmological constant problem \cite{Zeldovich:1967gd,Weinberg:1988cp}, lately there have been proposals where the cosmological constant term in the gravitational Lagrangian shows up as an integration constant, decoupled from the microphysics of vacuum energy \cite{Gallagher:2021tgx,Feng:2024rnh}. Notably, the cosmological constant term also appears as an integration constant during the reconstruction of $\Lambda$CDM-mimicking $f(Q)$ gravity for all the three possible connection branches. In fact, one encounters a similar situation while reconstructing the $\Lambda$CDM-mimicking $f(R)$ gravity \cite{Dunsby:2010wg}. 

In the context of modified gravity theories, a very pertinent question is whether one can achieve an exactly $\Lambda$CDM-like evolution with a vanishing cosmological constant term. If feasible, then one can attribute the observed late time cosmic evolution solely to additional \emph{dynamical} degrees of freedom coming the theory, and the cosmological constant issue simply never arises. Our results are as follows:

\begin{itemize}
\item \textbf{For the trivial connection { \boldmath$\Gamma_1$}}, this is not possible, as Eq.~\eqref{coeffs_rel_Gamma1} indicates that $\Lambda = 0$ implies $\Omega_{m0} = 0$.

\item \textbf{ For the connection { \boldmath$\Gamma_2$}}, it is possible to set $\Lambda = 0$, provided the remaining model parameters $c$ and $D$ satisfy the following bounds:
\be 
|c| < \frac{3}{2}\left(\frac{c_1}{1+\sqrt{1-c_1}}\right)\,, \qquad \frac{2|c|}{3c_1}\sqrt{1-c_1} \leq D < \frac{2|c|}{3c_1}\sqrt{1-c_1}\sqrt{1 + \frac{3}{2c^2(1-c_1)}}\,.
\ee 
Here, the first bound is obtained by combining the inequalities \eqref{c-D-bound} and \eqref{Lambda-bound} setting $\Lambda=0$. The second bound is obtained by combining the inequalities \eqref{Lambda-bound} and the requirement $0<\Omega_{m0}<1$, using the expression of $\Omega_{m0}$ from \eqref{coeffs_rel2_Gamma2} and setting $\Lambda=0$. 

\item  \textbf{For the connection  { \boldmath$\Gamma_3$}}, we were unable to derive an analytic form for the reconstructed $f(Q)$ that reproduces the full $\Lambda$CDM-like cosmic history. However, for the approximate reconstructed $f(Q)$ \eqref{reconfinal_Gamma3_approx}, valid during the matter-dominated epoch, we see that  setting $\Lambda = 0$ results in $C = 0$ (Eq.~\eqref{f-eq_approx}), which reduces the reconstructed $f(Q)$ \eqref{reconfinal_Gamma3_approx} to STEGR, $f(Q) \simeq Q$.
\end{itemize}

Conclusively, we can say that only the nontrivial connection branch $\Gamma_2$ allows us to achieve an exactly $\Lambda$CDM-like cosmic evolution without actually a cosmological constant term, thereby addressing the cosmological constant problem.

It is worth noting that we have specified the $\Lambda$CDM-like cosmic evolution as (Eq.\eqref{HLCDM})
\be
h_{\Lambda\rm CDM}(z)=\sqrt{\left[c_1(1+z)^3 + (1-c_1)\right]}\,, \qquad c_1=\frac{2}{3}(1+q_0)\,,
\ee 
instead of
\be
h_{\Lambda\rm CDM}(z)=\sqrt{\left[\Omega_{m0}(1+z)^3 + (1-\Omega_{m0})\right]}\,.
\ee 
While the identification $\Omega_{m0} = c_1 = \frac{2}{3}(1+q_0)$ holds for the actual General Relativistic $\Lambda$CDM model, it does not generally apply to alternative late-time models. In particular, for our reconstructed $f(Q)$ models, $\Omega_{m0}$ is expressed in terms of the other free model parameters and the present-day cosmographic parameters as follows:

\begin{itemize}

\item For {\boldmath $\Gamma_1$} (see \eqref{coeffs_rel_Gamma1}):
\be 
\Omega_{m0} = \frac{2\Lambda}{3H_0^2}\left(\frac{1+q_0}{1-2q_0}\right)\,.
\ee

\item For {\boldmath\textbf{$\Gamma_2$}} (see \eqref{coeffs_rel2_Gamma2}):
\be
\Omega_{m0} = \frac{2}{3}\lambda\left(\frac{1 + q_0}{1 - 2 q_0}\right) \pm \frac{2}{3}\sqrt{(1 + q_0)^2 \left(\frac{\lambda}{1-2q_0} - D\right)^2 -\frac{c^2}{3}(1 - 2q_0)}\,. 
\ee

\item For {\boldmath\textbf{$\Gamma_3$}}, $\Omega_{m0}$ is determined by solving the following equation(see \eqref{fx}):
\be 
f_0 = -\frac{6 c }{ x_0^2} + f_{Q0} (6 q_0 + \tilde{Q}_0)  -3 \Omega _{m0} \,,
\ee
the subscript `$0$' denoting present day values.
\end{itemize}

However, the converse is not always true. For example:
\begin{itemize}
\item For the case of { \boldmath$\Gamma_1$}, demanding that STEGR acts as a cosmological past attractor for the reconstructed $f(Q)$ ultimately leads us to the condition $\Omega_{m0}=c_1$, even though the theory starts deviating from STEGR at lower redshifts; see section \ref{subsec:freeparams_Gamma1}.
\item For the case of { \boldmath$\Gamma_2$}, employing the parameter choice of Eq.\eqref{coeffschoice-Gamma2}, we arrive at the 1-parameter reconstructed $f(Q)$ \eqref{reconsimple-Gamma2}, which also produces $\Omega_{m0}=c_1$, while the gravity theory is still different from STEGR. 
\end{itemize}

By combining CC+BAO+Pantheon+CMB data,  and employing a Gaussian process, one can obtain a model-independent estimate for $q_0$ typically as \cite[Table 5]{Mukherjee:2020ytg} $q_0=-0.647\pm0.069$. Similarly, combining CC+Type 1a Supernovae+BAO+Redshift Space Distorsion data and employing a Gaussian process one can obtain a model-independent estimate for $\Omega_{m0}$ typically as $0.224\pm0.066$ \cite{Ruiz-Zapatero:2022zpx}. The quantity $|c_1-\Omega_{m0}|/\Omega_{m0} \approx 5\%$ gives an estimate of the difference from the General Relativistic $\Lambda$CDM model according to a combination of different datasets. The difference between $c_1$ and $\Omega_{m0}$ holds the potential of possible amelioration of the Hubble tension \cite{Pedrotti:2024kpn}.  Our analysis reveals that the connection branch $\Gamma_2$ exhibits greater phenomenological potential than $\Gamma_1$, as it can maintain coincidence with STEGR at high redshifts without requiring $\Omega_{m0} = c_1$.

Our work paves the way for reconstructing $f(Q)$ theories from cosmological data for all three possible branches of symmetric teleparallel homogeneous and isotropic connections. By assuming no coupling between the matter Lagrangian and the connection, we simplify the equations by setting the hypermomentum to zero. However, in general, there is no inherent physical motivation to neglect hypermomentum within the teleparallel framework. We hope this work serves as a foundation for future efforts to reconstruct $f(Q)$ models, either analytically or numerically, in scenarios that include hypermomentum.

We must mention here that proving the existence of $\Lambda$CDM-mimicking modified gravity theories is not sufficient. It is crucial to explore ways to distinguish $\Lambda$CDM-mimicking models from the actual $\Lambda$CDM model. While the two scenarios may yield indistinguishable background evolution, differences will customarily arise in the perturbation evolution. This is due to the fact that the effective gravitational constant in general becomes a function of space-time in modified gravity theories. The change this brings into the perturbation evolution is apparent from the perturbation evolution equation (e.g. see \cite[Eq.7]{Anagnostopoulos:2021ydo} for the case of coincident gauge $f(Q)$ grvaity. Also see appendix \ref{app:distinguish}). Recently, a cosmographic approach was used to analyze perturbations in $\Lambda$CDM-mimicking $f(R)$ models, identifying distinctive signatures such as dispersion in the growth-index parameter \cite{MacDevette:2024wpg}. A similar analysis should be extended to $\Lambda$CDM-mimicking $f(Q)$ models across all the connection branches.

Finally, let us mention that in the era of precision cosmology it is an important task to asses the statistical likelihood of contesting models against the standard $\Lambda$CDM model. Statistical methodologies like the Akaike Information Criteria (AIC), Bayesian Information Criteria (BIC), Deviance Information Criteria (DIC) is being widely used for that purpose (see, e.g. \cite{Anagnostopoulos:2019miu,Anagnostopoulos:2021ydo,Atayde:2021pgb}). In general, additional parameters may enhance flexibility at the cost of reducing the constraining power of data. While our reconstructed $f(Q)$ models introduce additional parameters compared to the standard $\Lambda$CDM model, this is not without precedent in extended gravity models. For instance, the model $f(Q) = Q e^{\lambda \frac{Q_0}{Q}}$ proposed in \cite{Anagnostopoulos:2021ydo} features the same number of background parameters as the standard General Relativistic $\Lambda$CDM model. Notably, the reconstructed model for the connection branch $\Gamma_1$  has been shown in \cite{Atayde:2021pgb} to provide a better observational fit than $\Lambda$CDM even when RSD data is included. Hence, while additional parameters can pose a challenge, they may also improve the model's compatibility with data, highlighting the need for careful model selection analysis in future studies. Since we have been able to reconstruct the analytical form of the $\Lambda$CDM-mimicking $f(Q)$ for $\Gamma_2$, it is of particular interest to confront this model with data and assess its likelihood against the standard model.

Open questions remain, such as whether an analytic reconstruction for $\Gamma_3$ is possible and how the inclusion of hypermomentum might change the results. We leave them for future work.

\section*{Acknowledgments}

SC acknowledges funding support from the NSRF via the Program Management Unit for Human Resources and Institutional Development, Research and Innovation [grant number B13F670063]. JD acknowledges the support of IUCAA,
Pune (India) through the visiting associateship program. DG is a member of the GNFM working group of Italian INDAM, and also acknowledges financial support from the start-up funding plan of Jiangsu University of Science and Technology.

\appendix

\section{Distinguishing between the General Relativistic $\Lambda$CDM model and the $\Lambda$CDM-mimiking $f(Q)$ model for the case of $\Gamma_1$}\label{app:distinguish}

We have found that an $f(Q)$ gravitational model given by a linear function in the non-metricity scalar plus a cosmological constant term plus a term proportional to the square root of the abosolute value of the non-metricity scalar provides the same kinematics of a General Relativistic $\Lambda$CDM solution; specifically, the $\sqrt{-Q}$ term does not have any indentifying signature in the Friedmann and Raychaudhuri equations at the background level, but possibly have a signature in the evolution of the matter overdensity $\delta_m = \delta \rho_m/\rho_m$. The necessity of going to the perturbation level for distinguishing between models exhibiting the same background evolutions is well-known also in the context of similar works in the $f(R)$ realm \cite{Chakraborty:2021jku}. 

In the so-called coincidence gauge ($\Gamma_1$), the evolution of the matter overdensity in the linear approximation at sub-horizon scales is \cite[Eq.(7)]{Anagnostopoulos:2021ydo}
\beq
\frac{d^2 \delta_m}{da^2}+\left(\frac{dH}{H da} +\frac{3}{a}\right)\frac{d \delta_m}{da}-\frac{3 \Omega_{m0}}{2 H^2 a^5}\frac{\kappa_{\rm eff}}{\kappa}\delta_m=0\,.
\eeq
For our reconstructed $\Lambda$CDM-mimiking $f(Q)$ theory with (\ref{reconsimple-Gamma1}) and (\ref{HLCDM}), the latter equation specifies to
\beq
\frac{d^2 \delta_m}{da^2}+\left( 1-\frac{c_1}{2[c_1+(1-c_1)a^3]}\right)\frac{3}{a}\frac{d \delta_m}{da}-\frac{18 \Omega_{m0}}{H_0 \sqrt{c_1+(1-c_1)a^3}  [12 \sqrt{c_1+(1-c_1)a^3}H_0 -\sqrt{6}\beta a^{\frac{3}{2}}]a^2}\delta_m=0\,,
\eeq
which can be written more compactly as
\beq
\frac{d^2 \delta_m}{da^2}+\left(3-\frac{1+q_0}{a^3h^2(a)}\right)\frac{d \delta_m}{a da}-\frac{18 \Omega_{m0}}{H_0 h(a)  [12 H_0 h(a)  -\sqrt{6}\beta ]a^5}\delta_m=0\,.
\eeq

The evolution equation for the linear subhorizon matter overdensity\footnote{To be precise, evolution equation in the quasistatic approximation.} for modified gravity models like $f(R)$ or $f(Q)$ is a second order differential equation resembling that of a damped harmonic oscillator, with a purely kinematic damping term and a theory dependent \lq\lq force term". The particular theory enters the force term via $\kappa_{\rm eff}$. This theory-dependent force term is precisely the reason why the $\Lambda$CDM-mimiking $f(Q)$ model is expected to display its own specific signatures at the perturbation level that is different than that of the standard General Relativistic $\Lambda$CDM model as well as the $\Lambda$CDM-mimicking $f(R)$ models. Notice that, since the force term contains $\kappa_{\rm eff}=\frac{1}{f_Q}$ rather that $f(Q)$ itself, any possible presence of the cosmological constant in the reconstructed $f(Q)$ is oblivious to the perturbations.

\section{On the stability of the $\Lambda$CDM-like cosmic solution $j(z)=1$ in the reconstructed $f(Q)$ ($\Gamma_2$)}\label{app:stab_Gamma2}

In this section, we comment on the stability of the $\Lambda$CDM cosmological solution against small homogeneous and isotropic perturbations for the reconstructed $f(Q)$ for the case of $\Gamma_2$. To this goal, we first take a time derivative of the Raychaudhuri equation \eqref{raych-2}. After quite a few steps of calculations we arrive at the following equation
\be 
2(\ddot{H} + 9H\dot{H} + 9H^3)f_Q + 2(2\dot{H} + 3H^2)f_{QQ}\dot{Q} + 3(f - Qf_Q)H - f_{QQ}Q\dot{Q} = 0\,.
\ee
In deriving the above equation we have made use of the equation \eqref{Qdot-2}. Consider now a given solution $H(t)$ of the above equation, and take a homogeneous and isotropic perturbation around it: $H(t)\rightarrow H(t)+\delta H(t)$. We substitute it back in the above equation as well as in Eq.\eqref{Qdot-2}. Keeping terms of only linear order we get
\ba 
2 f_Q \ddot{\delta H} &+& (18H f_Q +4f_{QQ}\dot{Q})\dot{\delta H} + \left[18(1-q)H^2 f_Q + 12f_{QQ}H\dot{Q} + 3(f - Qf_Q)\right]\delta H \notag\\
&=& - f_{QQ}[2(1-2q)H^2 - Q]\dot{\delta Q} - [2(j-6q+2)H^3 f_{QQ} + 2(1-2q)H^2 \dot{Q}f_{QQQ} - (\dot{Q}+3HQ)f_{QQ} - Q\dot{Q}f_{QQQ}]\delta Q\,,\notag\\
&&\\
f_{QQ}\ddot{\delta Q} &+& (2\dot{Q}f_{QQQ} + 3Hf_{QQ})\dot{\delta Q} + [\dot{Q}^2 f_{QQQQ} + (\ddot{Q}+3H\dot{Q})f_{QQQ}]\delta Q = -3\dot{Q}f_{QQ}\delta H\,.
\ea 
For the particular 1-parameter reconstructed $f(Q)$ \eqref{reconsimple-Gamma2} that reproduces the $\Lambda$CDM-like background evolution \eqref{HLCDMfinal}, the above equations become
\ba 
&& \mathcal{A}(z)\ddot{\delta H} + \mathcal{B}(z)\dot{\delta H} + \mathcal{C}(z)\delta H = \mathcal{D}(z)\dot{\delta Q} + \mathcal{E}(z)\delta Q\,,
\label{eq-1}
\\
&& \ddot{\delta Q} + 3H\dot{\delta Q} = -3\dot{Q}\delta H\,,
\label{eq-2}
\ea 
where 
\begin{subequations}
\ba 
\mathcal{A}(z) &=& 2\left(D-\frac{2c}{3c_1}h_{\Lambda\rm CDM}(z)\right)\,,\\
\mathcal{B}(z) &=& \left[18h_{\Lambda\rm CDM}(z)D - \frac{4c}{3c_1}(7-2q_{\Lambda\rm CDM}(z))h_{\Lambda\rm CDM}^2(z)\right]H_0\,,\\
\mathcal{C}(z) &=& 18(1-q_{\Lambda\rm CDM}(z))H_{\Lambda\rm CDM}^2(z)\left(1-\frac{Q_{\Lambda\rm CDM}(z)}{6(1-c_1)H_0^2}\right) - \frac{2}{(1-c_1)H_0^2}H\dot{Q}_{\Lambda\rm CDM}(z) + 3\left(\Lambda + \frac{1}{12(1-c_1)H_0^2}Q_{\Lambda\rm CDM}^2\right)\,,\nonumber\\
&&\\
\mathcal{D}(z) &=& \frac{1}{6(1-c_1)H_0^2}[2(1-2q_{\Lambda\rm CDM}(z))H_{\Lambda\rm CDM}^2(z) - Q_{\Lambda\rm CDM}(z)]\,,\\
\mathcal{E}(z) &=& \frac{1}{6(1-c_1)H_0^2}[2(3-6q_{\Lambda\rm CDM}(z))H_{\Lambda\rm CDM}^3(z) - (\dot{Q}_{\Lambda\rm CDM}(z) + 3H_{\Lambda\rm CDM}(z)Q_{\Lambda\rm CDM}(z))]\,.
\ea 
\end{subequations}
In the above, the subscript `$\Lambda$CDM' implies that the respective dynamical quantities are calculated for $\Lambda$CDM-like background evolution \eqref{HLCDMfinal}, $c_1=\frac{2}{3}(1+q_0)$ according to Eq.\eqref{c1-q0} and $c$ is given by the equation \eqref{coeffschoice-Gamma2}.

We will not execute the full analysis here due to it's computational complexity, but we will chalk out the schematics of how one can proceed with it. Taking a time derivative of Eq.\eqref{eq1} and utilizing Eq.\eqref{eq-2}, one can get the following equation
\be 
\mathcal{A}\dddot{\delta H} + (\dot{\mathcal{A}}+\mathcal{B})\ddot{\delta H} + (\dot{\mathcal{B}} + \mathcal{C})\dot{\delta H} + (\mathcal{C}+3\mathcal{D}\dot{Q})\delta H = (\dot{\mathcal{D}}+\mathcal{E}-3\mathcal{D}H)\dot{\delta Q} + \dot{\mathcal{E}}\delta Q\,.\label{eq3}
\ee 
Eqs.\eqref{eq1} and \eqref{eq3} can be solved to obtain
\be
\delta Q = \delta Q (\delta H,\dot{\delta H},\ddot{\delta H},\dddot{\delta H})\,, \qquad \dot{\delta Q} = \dot{\delta Q}(\delta H,\dot{\delta H},\ddot{\delta H},\dddot{\delta H})\,.
\ee 
In particular, one can take another derivative of the expression for $\dot{\delta Q}$ to obtain
\be 
\ddot{\delta Q} = \ddot{\delta Q}(\delta H,\dot{\delta H},\ddot{\delta H},\dddot{\delta H},\ddddot{\delta H})\,.
\ee
One can then substitute the expressions for $\dot{\delta Q}$ and $\ddot{\delta Q}$ thus obtained to Eq.\eqref{eq-2} to obtain a 4-th order homogeneous ordinary differential equation for $\delta H$
\be 
(..)\ddddot{\delta H} + (..)\dddot{\delta H} + (..)\ddot{\delta H} + (..)\dot{\delta H} + (..)\delta H = 0\,,
\ee
with the coefficients depending on background related dynamical quantities. The signs of the roots of the characteristic polynomial of the above differential equation gives the stability of the $\Lambda$CDM cosmological solution with respect to small homogeneous isotropic perturbations within the reconstructed $f(Q)$ \eqref{reconsimple-Gamma2}. In particular, for the solution to be stable, all the roots must be negative.

\section{Some comments regarding the dynamical connection function ($\Gamma_2$)}\label{app:gamma_Gamma2}

For any generic cosmic history $H(z)$, in the context of the connection $\Gamma_2$, the inverse of the effective gravitational coupling relates to the redshift via the Hubble rate as (\ref{iii}), 
\beq
f_Q(z) = -c {\mathcal I(z)}+D\,,
\eeq
where we have defined 
\be 
{\mathcal I(z)} = \int \frac{(1+z)^2 dz}{h(z)}\,.
\ee 
Qualitatively, we can conclude from (\ref{iii}) that $f_Q(z)$ is a monotonic function of $z$ for any cosmic history of the universe (whether monotonically increasing or decreasing depends on the sign of the parameter $c$). The dynamical connection function for any generic cosmic history $H(z)$ can be expressed as \cite[Eq.(29)]{Yang:2024tkw}
\begin{eqnarray}
\label{gengammaz}
\gamma(z) = \frac{2h(z)H'(z)\left[c \int \frac{(1+z)^2 dz}{h(z)}-D \right]}{3c(1+z)^2}+\frac{2H(z)}{3}+\frac{\Omega_{m0}H_0}{c} = - \frac{2h(z)H'(z)f_Q(z)}{3c(1+z)^2}+\frac{2H(z)}{3}+\frac{\Omega_{m0}H_0}{c}\,,
\end{eqnarray}
or
\begin{eqnarray}
\label{robggf}
\frac{\gamma}{H_0} = \frac{2c_1(1+q)h^2(z)\left[c{\mathcal I(z)}-D \right]}{3c(h^2_{\rm \Lambda CDM}(z)-1+c_1)} + \frac{2h(z)}{3} + \frac{\Omega_{m0}}{c} = - \frac{2c_1(1+q(z))h^2(z)f_Q(z)}{3c(h^2_{\rm \Lambda CDM}(z)-1+c_1)} + \frac{2h(z)}{3} + \frac{\Omega_{m0}}{c}\,.
\end{eqnarray}

For the Hubble rate corresponding to a generic constant jerk parameter throughout the entire comic history (Eq.(\ref{generalconstj})), the relevant integral ${\mathcal I(z)}$ cannot be computed in a closed-form. We can nevertheless get a taste of the robustness of the reconstructed $\Lambda$CDM-mimicking $f(Q)$ within the range of astrophysical uncertainty for the observed jerk function by noticing that
\beq
H(z)=H_{\rm \Lambda CDM}(z)+ \frac{H_0}{3}  \left(h_{\rm \Lambda CDM}(z) +\frac{2(c_1-1)}{h_{\rm \Lambda CDM}(z)}\right) \ln(1+z) \varepsilon   + O(\varepsilon^2)\,,
\eeq
and consequently 
\begin{eqnarray}
{\mathcal I(z)}&:=& \int \frac{(1+z)^2 dz}{h(z)} =  \int \frac{(1+z)^2 dz}{h_{\rm \Lambda CDM}(z)} - \frac{\varepsilon }{3}\int (1+z)^2 \left( h_{\rm \Lambda CDM}(z) +\frac{2(c_1-1)}{h_{\rm \Lambda CDM}(z)}\right) \ln(1+z) dz + O(\varepsilon^2)\,, \nonumber\\
&=&  \frac{2h_{\rm \Lambda CDM}(z)}{3c_1}+\frac{2\varepsilon }{243c_1} \Big[30 (1-c_1)^{3/2}\arctan \left(\frac{h_{\rm \Lambda CDM}(z)}{\sqrt{1-c_1} }\right) \nonumber\\
&& \,\,\,- \left( 2[15(1-c_1)-h^2_{\rm \Lambda CDM}(z)]+9 [6(1-c_1)-h^2_{\rm \Lambda CDM}(z)] \ln \left( \frac{h^2_{\rm \Lambda CDM}(z)-1+c_1}{c_1}\right)^{1/3}  \right)h_{\rm \Lambda CDM}(z)    \Big]  + O(\varepsilon^2) \,. \nonumber\\
&& 
\end{eqnarray}
The deceleration parameter can be calculate similarly from Eq.(\ref{qconstj}). 

The case of the dynamical connection is therefore richer than that of the effective coupling, potentially providing a route for  discriminating between different non-metrical models. In fact, its increasing/decreasing properties appear  to be sensitive in a non-trivial manner to the algebraic sign of $\varepsilon$ because the former addendum is the product of the three functions $-2/(1+z)^3$ (negative and monotonically increasing), $f_Q(z)/c$ (negative and monotonically decreasing, at least in the limit $D \to 0$), and $(1+q(z))h^2(z)\equiv c_1k_1(1+z)^{k_1}+(1-c_1)k_2(1+z)^{k_2}$ (monotonically increasing with its algebraic sign depending on the value of $k_2$), and also the properties of the second addendum depend on the sign of the parameter $k_2$. Our  result (\ref{robggf}) constitutes the starting point for discriminating between different underlying models since the reconstructed dynamical functions from astrophysical data indeed exhibit different qualitative behaviors \cite{Yang:2024tkw}. Furthermore, the dynamical connection in turns determines the Riemann curvature entering the equation of geodesic deviations \cite{Beh:2021wva}, potentially bearing consequences on the propagation of gravitational wave signals \cite{Capozziello:2024vix}: identifying the  signature of the deperature from a $\Lambda$CDM-mimiking model, and specifically towards larger or smaller values of the jerk function, on their spectra is left for future investigations.

\bibliography{refs}

\begin{thebibliography}{10}

\bibitem{SupernovaSearchTeam:1998fmf}
Adam~G. Riess et~al.
\newblock {Observational evidence from supernovae for an accelerating universe
  and a cosmological constant}.
\newblock {\em Astron. J.}, 116:1009--1038, 1998.

\bibitem{SupernovaCosmologyProject:1998vns}
S.~Perlmutter et~al.
\newblock {Measurements of $\Omega$ and $\Lambda$ from 42 High Redshift
  Supernovae}.
\newblock {\em Astrophys. J.}, 517:565--586, 1999.

\bibitem{Weinberg:1988cp}
Steven Weinberg.
\newblock {The Cosmological Constant Problem}.
\newblock {\em Rev. Mod. Phys.}, 61:1--23, 1989.

\bibitem{Verde:2019ivm}
L.~Verde, T.~Treu, and A.~G. Riess.
\newblock {Tensions between the Early and the Late Universe}.
\newblock {\em Nature Astron.}, 3:891, 2019.

\bibitem{BeltranJimenez:2019esp}
Jose Beltr\'an~Jim\'enez, Lavinia Heisenberg, and Tomi~S. Koivisto.
\newblock {The Geometrical Trinity of Gravity}.
\newblock {\em Universe}, 5(7):173, 2019.

\bibitem{Aldrovandi:2015wfa}
R.~Aldrovandi and J.~G. Pereira.
\newblock {Teleparallelism: A New Way to Think the Gravitational Interaction}.
\newblock {\em Ciencia Hoje}, 55:32, 2015.

\bibitem{BeltranJimenez:2017tkd}
Jose Beltr\'an~Jim\'enez, Lavinia Heisenberg, and Tomi Koivisto.
\newblock {Coincident General Relativity}.
\newblock {\em Phys. Rev. D}, 98(4):044048, 2018.

\bibitem{Boehmer:2023fyl}
Christian~G. Boehmer and Erik Jensko.
\newblock {Modified gravity: A unified approach to metric-affine models}.
\newblock {\em J. Math. Phys.}, 64(8):082505, 2023.

\bibitem{Boehmer:2021aji}
Christian~G. Boehmer and Erik Jensko.
\newblock {Modified gravity: A unified approach}.
\newblock {\em Phys. Rev. D}, 104(2):024010, 2021.

\bibitem{Boehmer:2022wln}
Christian~G. Boehmer, Erik Jensko, and Ruth Lazkoz.
\newblock {Cosmological dynamical systems in modified gravity}.
\newblock {\em Eur. Phys. J. C}, 82(6):500, 2022.

\bibitem{Gomes:2023tur}
D\'ebora~Aguiar Gomes, Jose Beltr\'an~Jim\'enez, Alejandro~Jim\'enez Cano, and
  Tomi~S. Koivisto.
\newblock {Pathological Character of Modifications to Coincident General
  Relativity: Cosmological Strong Coupling and Ghosts in f(Q) Theories}.
\newblock {\em Phys. Rev. Lett.}, 132(14):141401, 2024.

\bibitem{Hu:2023gui}
Kun Hu, Makishi Yamakoshi, Taishi Katsuragawa, Shin'ichi Nojiri, and Taotao
  Qiu.
\newblock {Nonpropagating ghost in covariant f(Q) gravity}.
\newblock {\em Phys. Rev. D}, 108(12):124030, 2023.

\bibitem{BeltranJimenez:2020fvy}
Jose Beltr\'an~Jim\'enez, Alexey Golovnev, Tomi Koivisto, and Hardi Veerm\"ae.
\newblock {Minkowski space in $f(T)$ gravity}.
\newblock {\em Phys. Rev. D}, 103(2):024054, 2021.

\bibitem{Heisenberg:2023wgk}
Lavinia Heisenberg, Manuel Hohmann, and Simon Kuhn.
\newblock {Cosmological teleparallel perturbations}.
\newblock {\em JCAP}, 03:063, 2024.

\bibitem{Bello-Morales:2024vqk}
Antonio~G. Bello-Morales, Jose Beltr\'an~Jim\'enez, Alejandro Jim\'enez~Cano,
  Tomi~S. Koivisto, and Antonio~L. Maroto.
\newblock {A class of ghost-free theories in symmetric teleparallel geometry}.
\newblock {\em JHEP}, 12:146, 2024.

\bibitem{Anagnostopoulos:2021ydo}
Fotios~K. Anagnostopoulos, Spyros Basilakos, and Emmanuel~N. Saridakis.
\newblock {First evidence that non-metricity f(Q) gravity could challenge
  \ensuremath{\Lambda}CDM}.
\newblock {\em Phys. Lett. B}, 822:136634, 2021.

\bibitem{Atayde:2021pgb}
Lu\'\i{}s Atayde and Noemi Frusciante.
\newblock {Can $f(Q)$ gravity challenge $\Lambda$CDM?}
\newblock {\em Phys. Rev. D}, 104(6):064052, 2021.

\bibitem{Albuquerque:2022eac}
In\^es~S. Albuquerque and Noemi Frusciante.
\newblock {A designer approach to f(Q) gravity and cosmological implications}.
\newblock {\em Phys. Dark Univ.}, 35:100980, 2022.

\bibitem{Sahlu:2024pxk}
Shambel Sahlu, Alvaro de~la Cruz-Dombriz, and Amare Abebe.
\newblock {Structure growth in $f(Q)$ cosmology}.
\newblock {\em arXiv:2405.07361 [gr-qc]}, 5 2024.

\bibitem{Goncalves:2024sem}
Tiago~B. Gon\c{c}alves, Lu\'\i{}s Atayde, and Noemi Frusciante.
\newblock {Cosmological study of a symmetric teleparallel gravity model}.
\newblock {\em Phys. Rev. D}, 109(8):084003, 2024.

\bibitem{De:2022jvo}
Avik De and Tee-How Loo.
\newblock {On the viability of f(Q) gravity models}.
\newblock {\em Class. Quant. Grav.}, 40(11):115007, 2023.

\bibitem{Lazkoz:2019sjl}
Ruth Lazkoz, Francisco S.~N. Lobo, Mar\'\i{}a Ortiz-Ba\~nos, and Vincenzo
  Salzano.
\newblock {Observational constraints of $f(Q)$ gravity}.
\newblock {\em Phys. Rev. D}, 100(10):104027, 2019.

\bibitem{Ayuso:2020dcu}
Ismael Ayuso, Ruth Lazkoz, and Vincenzo Salzano.
\newblock {Observational constraints on cosmological solutions of $f(Q)$
  theories}.
\newblock {\em Phys. Rev. D}, 103(6):063505, 2021.

\bibitem{Frusciante:2021sio}
Noemi Frusciante.
\newblock {Signatures of $f(Q)$-gravity in cosmology}.
\newblock {\em Phys. Rev. D}, 103(4):044021, 2021.

\bibitem{Anagnostopoulos:2022gej}
Fotios~K. Anagnostopoulos, Viktor Gakis, Emmanuel~N. Saridakis, and Spyros
  Basilakos.
\newblock {New models and big bang nucleosynthesis constraints in f(Q)
  gravity}.
\newblock {\em Eur. Phys. J. C}, 83(1):58, 2023.

\bibitem{Planck:2018vyg}
N.~Aghanim et~al.
\newblock {Planck 2018 results. VI. Cosmological parameters}.
\newblock {\em Astron. Astrophys.}, 641:A6, 2020.
\newblock [Erratum: Astron.Astrophys. 652, C4 (2021)].

\bibitem{DESI:2023ytc}
A.~G. Adame et~al.
\newblock {The Early Data Release of the Dark Energy Spectroscopic Instrument}.
\newblock {\em Astron. J.}, 168(2):58, 2024.

\bibitem{Cortes:2024lgw}
Marina Cort\^es and Andrew~R. Liddle.
\newblock {Interpreting DESI's evidence for evolving dark energy}.
\newblock {\em JCAP}, 12:007, 2024.

\bibitem{Dunsby:2010wg}
Peter K.~S. Dunsby, Emilo Elizalde, Rituparno Goswami, Sergei Odintsov, and
  Diego~Saez Gomez.
\newblock {On the LCDM Universe in f(R) gravity}.
\newblock {\em Phys. Rev. D}, 82:023519, 2010.

\bibitem{He:2012rf}
Jian-hua He and Bin Wang.
\newblock {Revisiting $f(R)$ gravity models that reproduce $\Lambda$CDM
  expansion}.
\newblock {\em Phys. Rev. D}, 87(2):023508, 2013.

\bibitem{Fay:2007uy}
S.~Fay, S.~Nesseris, and L.~Perivolaropoulos.
\newblock {Can f(R) Modified Gravity Theories Mimic a LCDM Cosmology?}
\newblock {\em Phys. Rev. D}, 76:063504, 2007.

\bibitem{Elizalde:2010jx}
E.~Elizalde, R.~Myrzakulov, V.~V. Obukhov, and D.~Saez-Gomez.
\newblock {LambdaCDM epoch reconstruction from F(R,G) and modified Gauss-Bonnet
  gravities}.
\newblock {\em Class. Quant. Grav.}, 27:095007, 2010.

\bibitem{Gadbail:2022jco}
Gaurav~N. Gadbail, Sanjay Mandal, and P.~K. Sahoo.
\newblock {Reconstruction of \ensuremath{\Lambda}CDM universe in f(Q) gravity}.
\newblock {\em Phys. Lett. B}, 835:137509, 2022.

\bibitem{Hohmann:2021ast}
Manuel Hohmann.
\newblock {General covariant symmetric teleparallel cosmology}.
\newblock {\em Phys. Rev. D}, 104(12):124077, 2021.

\bibitem{Paliathanasis:2023hqq}
Andronikos Paliathanasis.
\newblock {The impact of the non-coincidence gauge on the dark energy dynamics
  in f(Q)-gravity}.
\newblock {\em Gen. Rel. Grav.}, 55(11):130, 2023.

\bibitem{Capozziello:2022wgl}
Salvatore Capozziello and Rocco D'Agostino.
\newblock {Model-independent reconstruction of f(Q) non-metric gravity}.
\newblock {\em Phys. Lett. B}, 832:137229, 2022.

\bibitem{Esposito:2021ect}
Fabrizio Esposito, Sante Carloni, Roberto Cianci, and Stefano Vignolo.
\newblock {Reconstructing isotropic and anisotropic f(Q) cosmologies}.
\newblock {\em Phys. Rev. D}, 105(8):084061, 2022.

\bibitem{Nojiri:2024zab}
Shin'ichi Nojiri and S.~D. Odintsov.
\newblock {Well-defined f(Q) gravity, reconstruction of FLRW spacetime and
  unification of inflation with dark energy epoch}.
\newblock {\em Phys. Dark Univ.}, 45:101538, 2024.

\bibitem{Yang:2024tkw}
Yuhang Yang, Xin Ren, Bo~Wang, Yi-Fu Cai, and Emmanuel~N. Saridakis.
\newblock {Data reconstruction of the dynamical connection function in f(Q)
  cosmology}.
\newblock {\em Mon. Not. Roy. Astron. Soc.}, 533(2):2232--2241, 2024.

\bibitem{naturej}
E.~R. {Harrison}.
\newblock {Observational tests in cosmology}.
\newblock {\em Nature}, 260(5552):591--592, April 1976.

\bibitem{Chakraborty:2022evc}
Saikat Chakraborty, Daniele Gregoris, and B.~Mishra.
\newblock {On the uniqueness of \ensuremath{\Lambda}CDM-like evolution for
  homogeneous and isotropic cosmology in General Relativity}.
\newblock {\em Phys. Lett. B}, 842:137962, 2023.

\bibitem{Sahni:2002fz}
Varun Sahni, Tarun~Deep Saini, Alexei~A. Starobinsky, and Ujjaini Alam.
\newblock {Statefinder: A New geometrical diagnostic of dark energy}.
\newblock {\em JETP Lett.}, 77:201--206, 2003.

\bibitem{Bernal:2016gxb}
Jose~Luis Bernal, Licia Verde, and Adam~G. Riess.
\newblock {The trouble with $H_0$}.
\newblock {\em JCAP}, 10:019, 2016.

\bibitem{Mukherjee:2020ytg}
Purba Mukherjee and Narayan Banerjee.
\newblock {Non-parametric reconstruction of the cosmological $jerk$ parameter}.
\newblock {\em Eur. Phys. J. C}, 81(1):36, 2021.

\bibitem{Jiang:2024xnu}
Jun-Qian Jiang, Davide Pedrotti, Simony~Santos da~Costa, and Sunny Vagnozzi.
\newblock {Nonparametric late-time expansion history reconstruction and
  implications for the Hubble tension in light of recent DESI and type Ia
  supernovae data}.
\newblock {\em Phys. Rev. D}, 110(12):123519, 2024.

\bibitem{Pogosian:2021mcs}
Levon Pogosian, Marco Raveri, Kazuya Koyama, Matteo Martinelli, Alessandra
  Silvestri, Gong-Bo Zhao, Jian Li, Simone Peirone, and Alex Zucca.
\newblock {Imprints of cosmological tensions in reconstructed gravity}.
\newblock {\em Nature Astron.}, 6(12):1484--1490, 2022.

\bibitem{Bardeen:1980kt}
James~M. Bardeen.
\newblock {Gauge Invariant Cosmological Perturbations}.
\newblock {\em Phys. Rev. D}, 22:1882--1905, 1980.

\bibitem{Luongo:2013rba}
Orlando Luongo.
\newblock {Dark energy from a positive jerk parameter}.
\newblock {\em Mod. Phys. Lett. A}, 28:1350080, 2013.

\bibitem{Amirhashchi:2018vmy}
Hassan Amirhashchi and Soroush Amirhashchi.
\newblock {Recovering $\Lambda $CDM model from a cosmographic study}.
\newblock {\em Gen. Rel. Grav.}, 52(2):13, 2020.

\bibitem{Ortiz-Banos:2021jgg}
Mar\'\i{}a Ortiz-Ba\~nos, Mariam Bouhmadi-L\'opez, Ruth Lazkoz, and Vincenzo
  Salzano.
\newblock {${\Lambda}$CDM suitably embedded in f(R) with a non-minimal coupling
  to matter}.
\newblock {\em Eur. Phys. J. C}, 81(3):237, 2021.

\bibitem{Choudhury:2019zod}
Shibendu~Gupta Choudhury, Ananda Dasgupta, and Narayan Banerjee.
\newblock {Reconstruction of $f(R)$ gravity models for an accelerated universe
  using the Raychaudhuri equation}.
\newblock {\em Mon. Not. Roy. Astron. Soc.}, 485(4):5693--5699, 2019.

\bibitem{Gadbail:2023mvu}
Gaurav~N. Gadbail, Avik De, and P.~K. Sahoo.
\newblock {Cosmological reconstruction and $\Lambda $CDM universe in $f(Q,\,C)$
  gravity}.
\newblock {\em Eur. Phys. J. C}, 83(12):1099, 2023.

\bibitem{Carloni:2010ph}
Sante Carloni, Rituparno Goswami, and Peter K.~S. Dunsby.
\newblock {A new approach to reconstruction methods in $f(R)$ gravity}.
\newblock {\em Class. Quant. Grav.}, 29:135012, 2012.

\bibitem{Dunajski:2008tg}
Maciej Dunajski and Gary Gibbons.
\newblock {Cosmic Jerk, Snap and Beyond}.
\newblock {\em Class. Quant. Grav.}, 25:235012, 2008.

\bibitem{Song:2006ej}
Yong-Seon Song, Wayne Hu, and Ignacy Sawicki.
\newblock {The Large Scale Structure of f(R) Gravity}.
\newblock {\em Phys. Rev. D}, 75:044004, 2007.

\bibitem{Pogosian:2007sw}
Levon Pogosian and Alessandra Silvestri.
\newblock {The pattern of growth in viable f(R) cosmologies}.
\newblock {\em Phys. Rev. D}, 77:023503, 2008.
\newblock [Erratum: Phys.Rev.D 81, 049901 (2010)].

\bibitem{Kumar:2016bzd}
Rohin Kumar.
\newblock {A New Class of Cosmologically `Viable' $f(R)$ Models}.
\newblock {\em arXiv: 1611.03728 [gr-qc]}, 11 2016.

\bibitem{Chakraborty:2021jku}
Saikat Chakraborty, Kelly MacDevette, and Peter Dunsby.
\newblock {A model independent approach to the study of $f(R)$ cosmologies with
  expansion histories close to $\Lambda$CDM}.
\newblock {\em Phys. Rev. D}, 103(12):124040, 2021.

\bibitem{Zhai:2013fxa}
Zhong-Xu Zhai, Ming-Jian Zhang, Zhi-Song Zhang, Xian-Ming Liu, and Tong-Jie
  Zhang.
\newblock {Reconstruction and constraining of the jerk parameter from OHD and
  SNe Ia observations}.
\newblock {\em Phys. Lett. B}, 727:8--20, 2013.

\bibitem{Pedrotti:2024kpn}
Davide Pedrotti, Jun-Qian Jiang, Luis~A. Escamilla, Simony~Santos da~Costa, and
  Sunny Vagnozzi.
\newblock {Multidimensionality of the Hubble tension: The roles of
  \ensuremath{\Omega}m and \ensuremath{\omega}c}.
\newblock {\em Phys. Rev. D}, 111(2):023506, 2025.

\bibitem{1970PhT....23b..34S}
A.~R. {Sandage}.
\newblock {Cosmology: a search for two numbers.}
\newblock {\em Physics Today}, 23(2):34--41, January 1970.

\bibitem{Hohmann:2019fvf}
Manuel Hohmann.
\newblock {Metric-affine Geometries With Spherical Symmetry}.
\newblock {\em Symmetry}, 12(3):453, 2020.

\bibitem{Zhao:2021zab}
Dehao Zhao.
\newblock {Covariant formulation of f(Q) theory}.
\newblock {\em Eur. Phys. J. C}, 82(4):303, 2022.

\bibitem{Jarv:2018bgs}
Laur J\"arv, Mihkel R\"unkla, Margus Saal, and Ott Vilson.
\newblock {Nonmetricity formulation of general relativity and its scalar-tensor
  extension}.
\newblock {\em Phys. Rev. D}, 97(12):124025, 2018.

\bibitem{Guzman:2024cwa}
Maria-Jose Guzman, Laur J\"arv, and Laxmipriya Pati.
\newblock {Exploring the stability of f(Q) cosmology near general relativity
  limit with different connections}.
\newblock {\em Phys. Rev. D}, 110(12):124013, 2024.

\bibitem{BeltranJimenez:2019tme}
Jose Beltr\'an~Jim\'enez, Lavinia Heisenberg, Tomi~Sebastian Koivisto, and
  Simon Pekar.
\newblock {Cosmology in $f(Q)$ geometry}.
\newblock {\em Phys. Rev. D}, 101(10):103507, 2020.

\bibitem{Ray:2005ia}
Saibal Ray and Utpal Mukhopadhyay.
\newblock {Dark energy models with time-dependent gravitational constant}.
\newblock {\em Int. J. Mod. Phys. D}, 16:1791--1802, 2007.

\bibitem{Jamil:2009sq}
Mubasher Jamil, Emmanuel~N. Saridakis, and M.~R. Setare.
\newblock {Holographic dark energy with varying gravitational constant}.
\newblock {\em Phys. Lett. B}, 679:172--176, 2009.

\bibitem{Capozziello:2019cav}
Salvatore Capozziello, Rocco D'Agostino, and Orlando Luongo.
\newblock {Extended Gravity Cosmography}.
\newblock {\em Int. J. Mod. Phys. D}, 28(10):1930016, 2019.

\bibitem{Boehmer:2014vea}
Christian~G. Boehmer and Nyein Chan.
\newblock {\em {Dynamical systems in cosmology.}}
\newblock 2017.

\bibitem{Barrow:1983rx}
John~D. Barrow and A.~C. Ottewill.
\newblock {The Stability of General Relativistic Cosmological Theory}.
\newblock {\em J. Phys. A}, 16:2757, 1983.

\bibitem{delaCruz-Dombriz:2011oii}
Alvaro de~la Cruz-Dombriz and Diego Saez-Gomez.
\newblock {On the stability of the cosmological solutions in $f(R,G)$ gravity}.
\newblock {\em Class. Quant. Grav.}, 29:245014, 2012.

\bibitem{Wang:2024eai}
Qingqing Wang, Xin Ren, Yi-Fu Cai, Wentao Luo, and Emmanuel~N. Saridakis.
\newblock {Observational Test of f(Q) Gravity with Weak Gravitational Lensing}.
\newblock {\em Astrophys. J.}, 974(1):7, 2024.

\bibitem{Sahlu:2022bgy}
Shambel Sahlu and Endalkachew Tsegaye.
\newblock {Linear Cosmological perturbations in $f(Q)$ Gravity}.
\newblock {\em arXiv:2206.02517 [gr-qc]}, 6 2022.

\bibitem{Zeldovich:1967gd}
Y.~B. Zeldovich.
\newblock {Cosmological Constant and Elementary Particles}.
\newblock {\em JETP Lett.}, 6:316, 1967.

\bibitem{Gallagher:2021tgx}
Priidik Gallagher and Tomi Koivisto.
\newblock {The \ensuremath{\Lambda} and the CDM as Integration Constants}.
\newblock {\em Symmetry}, 13(11):2076, 2021.

\bibitem{Feng:2024rnh}
Justin~C. Feng and Pisin Chen.
\newblock {Cosmological constant as an integration constant}.
\newblock {\em Eur. Phys. J. C}, 84(12):1331, 2024.

\bibitem{Ruiz-Zapatero:2022zpx}
Jaime Ruiz-Zapatero, Carlos Garc\'\i{}a-Garc\'\i{}a, David Alonso, Pedro~G.
  Ferreira, and Richard D.~P. Grumitt.
\newblock {Model-independent constraints on \ensuremath{\Omega}m and H(z) from
  the link between geometry and growth}.
\newblock {\em Mon. Not. Roy. Astron. Soc.}, 512(2):1967--1984, 2022.

\bibitem{MacDevette:2024wpg}
Kelly MacDevette, Jess Worsley, Peter Dunsby, and Saikat Chakraborty.
\newblock {A model independent approach to the study of structure growth in
  $f(R)$ gravity}.
\newblock {\em arXiv:2408.03998 [gr-qc]}, 8 2024.

\bibitem{Anagnostopoulos:2019miu}
Fotios~K. Anagnostopoulos, Spyros Basilakos, and Emmanuel~N. Saridakis.
\newblock {Bayesian analysis of $f(T)$ gravity using $f\sigma_8$ data}.
\newblock {\em Phys. Rev. D}, 100(8):083517, 2019.

\bibitem{Beh:2021wva}
Jing-Theng Beh, Tee-how Loo, and Avik De.
\newblock {Geodesic deviation equation in f(Q)-gravity}.
\newblock {\em Chin. J. Phys.}, 77:1551--1560, 2022.

\bibitem{Capozziello:2024vix}
Salvatore Capozziello, Maurizio Capriolo, and Shin'ichi Nojiri.
\newblock {Gravitational waves in f(Q) non-metric gravity via geodesic
  deviation}.
\newblock {\em Phys. Lett. B}, 850:138510, 2024.

\end{thebibliography}
\bibliographystyle{unsrt}

\end{document}